%% file: main.tex
\documentclass[12pt]{iopart}

\usepackage{citesort}
\usepackage{url}
\usepackage[a4paper, top=25mm, bottom=25mm, left=20mm, right=20mm]{geometry}

\usepackage{lineno}
\usepackage{graphicx}

\usepackage{multirow}

\begin{document}
\title[]{Scalable Multi-Task Learning for Particle Collision Event Reconstruction with Heterogeneous Graph Neural Networks} 

\author{William Sutcliffe$^1$, Marta Calvi$^{2,3}$, Simone Capelli$^{3}$, Jonas Eschle$^5$, Julián García Pardiñas$^4$, Abhijit Mathad$^5$, Azusa Uzuki$^1$ and Nicola Serra$^1$}
\address{$^1$ Department of Physics, University of Z{\"u}rich, Winterthurerstrasse 190, Z{\"u}rich, 8057, Switzerland.}
\address{$^2$ Dipartimento di Fisica “G. Occhialini”, Universit\`{a} di Milano Bicocca, Piazza della Scienza 3, Milano, 20126, Italy}
\address{$^3$ INFN Sezione di Milano-Bicocca, Piazza della Scienza 3, Milano, 20126, Italy}
\address{$^4$ Laboratory for Nuclear Science. Massachusetts Institute of Technology (MIT), 77 Massachusetts Ave, Cambridge, MA 02139, USA.}
\address{$^5$ Experimental Physics Department, European Organization for Nuclear Research
(CERN), Espl. des Particules 1, Meyrin, 1211, Switzerland.}

\ead{william.sutcliffe@cern.ch}

\begin{abstract}
The growing luminosity frontier at the Large Hadron Collider is challenging the reconstruction and analysis of particle collision events. Increased particle multiplicities are straining latency and storage requirements at the data acquisition stage, while new complications are emerging, including higher background levels and more frequent particle vertex misassociations. This in turn necessitates the development of more holistic and scalable reconstruction methods that take advantage of recent advances in machine learning. We propose a novel Heterogeneous Graph Neural Network (HGNN) architecture featuring unique representations for diverse particle collision relationships and integrated graph pruning layers for scalability. Trained with a multi-task paradigm in an environment mimicking the LHCb experiment, this HGNN significantly improves the beauty hadron reconstruction performance. Notably, it concurrently performs particle vertex association and graph pruning within a single framework. We quantify the reconstruction and pruning performance, demonstrate enhanced inference time scaling with event complexity, and mitigate potential performance loss using a weighted message passing scheme.
\end{abstract}

\section{Introduction}
\label{sec:intro}
\input{introduction}

\section{Related work}
\label{sec:recent_work}
\input{relatedwork}

\section{Methodology}
\label{sec:methodology}
\input{methodology}

\section{Results}
\label{sec:results}
\input{results}

\section{Discussion}
\label{sec:discussion}

\input{discussion}

\section{Future work}
\label{sec:future}
\input{future_work}

\section{Conclusion}
\label{sec:conclusion}
\input{conclusion}

\input{acknowledgements}
\input{declaration}
\input{data_availability}

\appendix
\input{appendix}

\section*{References}
\bibliography{iopart-num}

\end{document}

%% file: introduction.tex
Over the past two decades, the field of neural networks has witnessed rapid advancements, driving breakthroughs
across diverse domains such as natural language processing, image analysis, and scientific computation. Architectures ranging from convolutional neural networks~\cite{NIPS2012_c399862d} to more recent transformer~\cite{DBLP:journals/corr/VaswaniSPUJGKP17} models have played pivotal roles in these developments. Meanwhile, graph neural networks~\cite{DBLP:journals/corr/abs-1901-00596} (GNNs) have emerged as a powerful
tool for representing complex datasets of variable size that lack explicit spatial or sequential structures, effectively modeling interactions among multiple entities and their interconnections.
Building on this success, heterogeneous GNNs~\cite{yang2023simpleefficientheterogeneousgraph} (HGNNs) extend conventional GNNs by incorporating multiple types of nodes and edges, enabling richer and more context-specific representations for complex, multi-relational data. HGNNs find extensive applications in areas such as recommendation systems (modeling users, items, and interactions)~\cite{shi2020heterogeneousgraphneuralnetwork}, bioinformatics (gene-disease-drug networks)~\cite{zhang2024heterogeneouscausalmetapathgraph}, and natural language processing tasks such as cross-lingual text classification~\cite{wang-etal-2021-cross-lingual}.

In parallel, multi-task learning has gained traction as an approach that simultaneously optimises several
related objectives, allowing models to share representations and improve generalization across tasks~\cite{DBLP:journals/corr/abs-2009-09796}.
By leveraging the synergies among tasks, multi-objective learning can not only enhance performance but also provide a more holistic understanding of underlying data representations.

Concurrently, in particle physics recent trends have shown an increase in the application of machine learning (ML) across several domains including simulation, detector reconstruction and particle classification~\cite{Radovic:2018dip,Bourilkov:2019yoi}. HGNNs and multi-task learning are particularly suited to particle collider experiments given that particle collision events are inherently heterogeneous involving several object types and their relationships with one another. However, despite their potential, HGNNs have rarely been applied in particle physics~\cite{Huang:2023ssr,Caillou:2024dcn}, where most applications rely on homogeneous GNN architectures~\cite{Shlomi:2020gdn}. Unifying HGNNs and multi-task learning allows several reconstruction tasks to be performed in parallel. 

The LHCb experiment~\cite{LHCb:2008vvz} at CERN’s Large Hadron Collider focused on high-precision studies of beauty ($b$) and charm ($c$) quarks. Its main goal is to test the Standard Model (SM) through measurements of CP violation, rare beauty-hadron decays, and flavor-changing neutral current processes that are highly sensitive to new physics beyond the SM. During Runs 1 and 2 (2010-2018), LHCb recorded $9\,\mathrm{fb}^{-1}$ of proton-proton collision data at a luminosity of $4\times 10^{32}\,\mathrm{cm}^{-2}\,\mathrm{s}^{-1}$. A recent upgrade (Upgrade I) for Runs 3 and 4 increased the luminosity to $2\times 10^{33}\,\mathrm{cm}^{-2}\,\mathrm{s}^{-1}$, resulting in an average of five collisions per event and a charged particle track multiplicity of approximately 150~\cite{LHCb:2023hlw}. Looking ahead, the recently approved LHCb Upgrade II at the High-Luminosity LHC is expected to boost the luminosity tenfold, yielding on average 50 collisions per event~\cite{LHCb:2018roe} and a track multiplicity of around 1,000 particles.

The increasing collision event complexity at the LHCb experiment presents significant challenges for data acquisition and measurement precision. While LHCb's Upgrade I trigger framework~\cite{Aaij:2019uij} has already transitioned to partial event storage, moving beyond the traditional approach of storing full events that was feasible when the disk space was less constrained~\cite{LHCb:2018zdd}, the challenges will intensify under Upgrade II conditions. The anticipated increase in particle multiplicity and the common occurrence of multiple heavy-hadron decays within single events will further strain storage resources.  Consequently, refining strategies to systematically identify and preserve the most valuable event components remains an important objective, especially when considering the diverse heavy-hadron species and decays per species, whose combinatorics amount to tens of thousands of different possible decays of interest. While exclusive selections are advantageous for storage, capturing sufficient information for studies requiring context from the underlying event (e.g., for flavour tagging or measurements of semileptonic decays) presents an ongoing challenge. Developing techniques that can selectively preserve this broader event information, when needed for specific physics goals, while adhering to stringent latency constraints ($\mathcal{O}$(100ms) per event and CPU core) and storage limitations ($\mathcal{O}$(10PB) per year), will be beneficial for maximising the physics potential of LHCb.

Another critical issue emerging in the context of increased luminosity conditions at the LHCb is primary vertex (PV) misassociation. PVs correspond to proton-proton interaction points, and under high-luminosity conditions, multiple such interactions can occur in a single event. PV misassociation arises when tracks or decay products from overlapping proton-proton collisions are incorrectly attributed to a PV. This can severely degrade the PV resolution and bias the measurement of key observables, such as the beauty-hadron decay flight distance and direction, ultimately affecting the precision of determinations of time-dependent CP violation~\cite{LHCb:2011aa} and measurements of decays with missing energy~\cite{Lbnat2015,LHCb:2024jll}. Therefore, addressing PV misassociation is paramount to maintaining the high-precision performance required for LHCb’s physics program in the high-luminosity era, prompting the need for innovative reconstruction algorithms and advanced machine learning techniques.

Recently, we proposed the Deep Full Event Interpretation (DFEI) algorithm~\cite{GarciaPardinas:2023pmx}, which employs GNNs to perform a multi-stage inclusive reconstruction of beauty hadrons in LHCb collision events. Despite its potential, the computational cost and scalability of its multi-stage approach pose significant challenges with regard to meeting the low-latency requirements of a real-time trigger environment. Furthermore, the algorithm did not address the pressing issue of PV misassociation. Building on the success of our earlier work, we propose an HGNN architecture with integrated graph pruning for scalability, which is trained with multiple objectives to perform beauty hadron reconstruction and PV association within a unified framework.

%% file: relatedwork.tex
Particle decay reconstruction typically follows an exclusive strategy in which final‑state particles are successively combined into intermediate structures to form complete decay chains for specific topologies. This process proceeds sequentially by combining final-state particles into higher-level structures based on predefined decay modes and kinematic constraints, such as invariant mass thresholds or the conservation of momentum and energy. At each level, particle identification techniques and machine learning classification algorithms can be applied to improve reconstruction accuracy and resolve ambiguities.

In addition to traditional methods that target individual decay channels, intermediate strategies exist that bridge these and the fully holistic reconstruction we propose. Key examples include the LHCb topological trigger~\cite{BBDTTopo,Schulte:2023gtt} and the Belle II tag-side reconstruction algorithm known as Full Event Interpretation (FEI)~\cite{Keck:2018lcd}.  The LHCb topological trigger identifies beauty-hadron decays based on predefined characteristic topologies, which rely on multivariate classifiers trained on an ensemble of decay modes, making their selection inherently guided by predefined exclusive states. Similarly, FEI performs a hierarchical reconstruction of a large number of beauty-hadron decay chains with a dedicated multivariate classifier for each unique particle decay.

More recently, several efforts have been made towards a fully inclusive reconstruction of beauty-hadron decays with GNNs at LHCb and Belle II~\cite{Kahn:2022njt,Abumusabh:2025tti,GarciaPardinas:2023pmx}. These developments were prompted by Kahn {\it et al.}~\cite{Kahn:2022njt} with the introduction of a novel edge classification target for hierarchical decay chains known as the lowest common ancestor generations (LCAG) matrix. This compact representation enables one to learn the hierarchical structure of a decay solely from its final-state particles. For each edge relation between final-state particles, a multi-class label is used, which signifies the generational class of the shared ancestor. Kahn {\it et al.} further benchmarked their GNN-based approach against transformer architectures for LCAG reconstruction, demonstrating significantly better performance of GNNs for this hierarchical task.

In our previous publication~\cite{GarciaPardinas:2023pmx}, the Deep Full Event Interpretation (DFEI), we expanded on this work demonstrating the inclusive reconstruction of beauty-hadron decays with GNNs within the hadronic environment of LHCb, which is complicated by a large number of background particles. To overcome these difficulties, we employed a multi-stage approach. First, a node-pruning GNN filters out background nodes based on kinematic and topological features. Next, an edge-pruning GNN removes unlikely associations and simplifies the event graph. Finally, a GNN processes the remaining graph and performs the edge classification of LCAG values, enabling the separation and hierarchical reconstruction of multiple possible beauty-hadron decay chains in each event.

Beyond particle decay reconstruction, GNNs have seen significant adoption in particle physics~\cite{Shlomi:2020gdn}. For charged-particle tracking, GNNs are employed to connect detector hits (nodes) into particle trajectories by classifying potential track segments (edges)~\cite{Elabd:2021lgo}. GNNs have demonstrated improved performance for flavor tagging of beauty-hadron events at Belle II with GFlaT~\cite{Belle-II:2024lwr}. GNNs are also increasingly used for jet classification and reconstruction tasks, representing jets as point clouds or graphs of constituent particles to distinguish between different originating particles and to better reconstruct the kinematic quantities of the jet~\cite{Qu:2019gqs,ATLAS:2022rkn,Guo:2020vvt}. Furthermore, GNNs are employed for particle flow (PF) algorithms, which aim to provide an end-to-end ML approach that combines information from different subdetectors to reconstruct a complete list of particles~\cite{Pata:2021oez}. While most applications use homogeneous GNNs, HGNNs have been applied to improve hadronic $\tau$ lepton identification by treating tracks and energy clusters as distinct node types within a jet graph~\cite{Huang:2023ssr} and in novel designs for track reconstruction that explicitly account for different detector sensor types (e.g., pixel vs. strip hits)~\cite{Caillou:2024dcn}. 

To manage computational costs and focus on relevant relations, two main strategies are employed: graph pruning and dynamic graph construction. Graph pruning typically starts with a larger, often geometrically constrained graph, and then removes edges deemed unlikely to represent true physical connections. This is common in tracking pipelines (e.g. ExaTrkX), where initial filtering steps or GNN-based edge classifiers prune the graph significantly before track finding~\cite{ExaTrkX:2021abe,Liu:2023siw}. Usually thresholding the output scores of edge-classifying GNNs serves as the pruning mechanism. Alternatively, dynamic graph construction methods can adapt the graph connectivity during the learning process. Techniques such as k-nearest neighbors (k-NN) applied in a learned latent space allow the graph structure to evolve, connecting nodes based on their learned representations. This is exemplified by architectures such as EdgeConv (used in ParticleNet for jet tagging)~\cite{Qu:2019gqs} and GravNet (used for calorimeter clustering)~\cite{Qasim:2019otl}, which dynamically define edges. Both pruning and dynamic construction aim to improve the overall performance and scalability of GNNs.

Finally, multi-task learning (MTL), the paradigm we adopt, has become increasingly relevant in particle physics, where complex analysis often involves inferring multiple correlated properties or performing hierarchical reconstruction steps. For instance, in jet physics, ``foundation models" are being developed using MTL principles, pre-training on large datasets for self-supervised tasks such as jet generation or masked particle prediction, and then fine-tuning for various downstream tasks such as jet tagging or property prediction, with the aim of creating a universal jet representation~\cite{Birk:2024knn,Golling:2024abg,Mikuni:2025tar}. In neutrino physics event reconstruction, NuGraph2 applied MTL with GNNs to simultaneously perform tasks such as detector hit classification and semantic segmentation (assigning hits to electron or muon particle types)~\cite{Aurisano:2024uvd}. The concept also extends to using auxiliary tasks to aid a primary goal, such as predicting track origins alongside jet flavour in ATLAS's GNN jet tagger (GN1)~\cite{ATLAS:2022rkn}. Beyond particle physics, MTL is also used in related fields such as fusion energy for real-time plasma equilibrium reconstruction, simultaneously predicting multiple plasma parameters and profile distributions~\cite{Fusion}.

%% file: methodology.tex
\subsection{Graph Neural Networks}
\label{sec_gnns}

GNNs encompass a diverse range of architectures, each tailored to specific tasks and graph structures. Key approaches include Graph Convolutional Networks (GCNs)~\cite{kipf2017gcn}, which extend traditional convolutions to graphs by aggregating and transforming features from connected nodes, and Graph Attention Networks (GATs)~\cite{velickovic2018graph}, which use attention mechanisms to weight node contributions. Message Passing Neural Networks (MPNNs)~\cite{gilmer2017neural} further unify these methods under a general framework where interconnected nodes iteratively exchange and update messages to learn graph representations.

In this paper we build upon the GNN introduced by Battaglia {\it et al.}~\cite{battaglia2018relationalinductivebiasesdeep}, which provides a more comprehensive and versatile framework for MPNNs learning representations at multiple levels, including nodes, edges, and globally, making it well suited to capture the hierarchical and relational information inherent in particle collisions. Equation~\ref{eq:GNeq1} summarises the update equations for the GNN, which consist of edge, node and graph updates. 
\begin{equation}
\begin{split}
e'^k = \phi^e(e^k, v^{r_{k}}, v^{s_{k}}, u) \qquad & \bar{e}'^i = \rho^{e\rightarrow v}(\{ E'^{i} \})\\
v'^i = \phi^v(\bar{e}'^i, v^i,  u) \qquad & \bar{e}' = \rho^{e\rightarrow u}(\{ E'  \})\\
u' = \phi^u(\bar{e}', \bar{v}', u)  \qquad & \bar{v}' = \rho^{v\rightarrow u}(\{ V' \})\\
\end{split}
\label{eq:GNeq1}
\end{equation}
The edge update function $\phi^e$ is a learnable multi-layer-perceptron (MLP), that takes as input the existing edge representation $e^k$ for edge $k$, adjoining receiver and sender node representations ($v^{r_{k}}$ and $v^{s_{k}}$) and a global representation $u$. The subsequent node update with MLP $\phi^v$ for node $i$, takes as input the node $v^i$ representation, the edge-to-node aggregation $\bar{e}'^i$ and $u$. Here, $\bar{e}'^i$ aggregates edge representations for the set of edges $E'^{i}$ that node $i$ receives using an aggregation function $\rho^{e\rightarrow v}$~\footnote{For example, if node $1$ receives edges from nodes $2$ and $3$, 
the incoming edges $(2\!\to\!1)$ and $(3\!\to\!1)$ are first updated as 
$e'^{21} = \phi^e(e^{21}, v^1, v^2, u)$ and 
$e'^{31} = \phi^e(e^{31}, v^1, v^3, u)$. 
These are then aggregated, for instance 
$\bar{e}'^{1} = e'^{21} + e'^{31}$ for sum pooling, 
and the node representation is updated as 
$v'^1 = \phi^v(v^1, \bar{e}'^{\,1}, u)$.}. Possible aggregation functions include non-parametric functions such as sum, mean, and max pooling. Finally, sets of edges and nodes, $E'$ and $V'$, are aggregated globally to give $\bar{e}'$ and $\bar{v}'$ using the aggregation functions $\rho^{e\rightarrow u}$ and $\rho^{v\rightarrow u}$. The global update $u'$ with MLP $\phi^u$ takes as input global aggregations of edges ($\bar{e}'$) and nodes ($\bar{v}'$) and the existing global representation, $u$.

\subsection{HGNN layer for particle reconstruction and pruning}
\label{sec:HGNN}

Unlike the uniform approach of homogeneous GNNs, HGNNs improve upon homogeneous GNNs by natively handling multiple node and edge types with type-specific representations and update functions. This inherent flexibility makes HGNNs better suited for modeling diverse systems such as physical interactions. Although homogeneous GNNs can mimic this heterogeneity using techniques such as one-hot encoding and padding, they lack the inherent inductive bias of HGNNs, often leading to suboptimal representations and making it difficult to set distinct learning objectives for different entity types.

\begin{figure}[!h]
\begin{center}
\hspace{-0.5cm}\includegraphics[width=9cm]{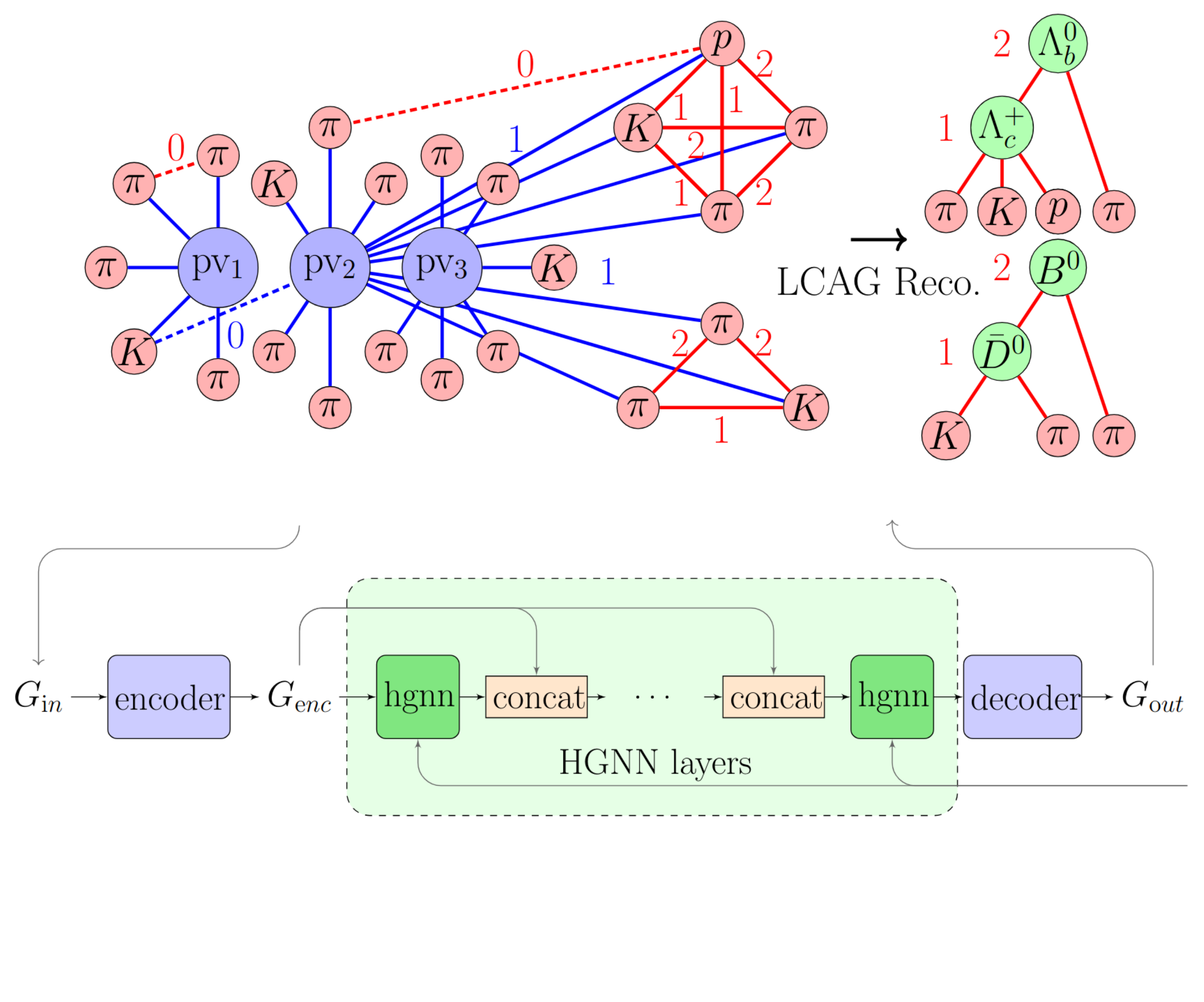}\put(-250,200){(a)}\includegraphics[width=8cm]{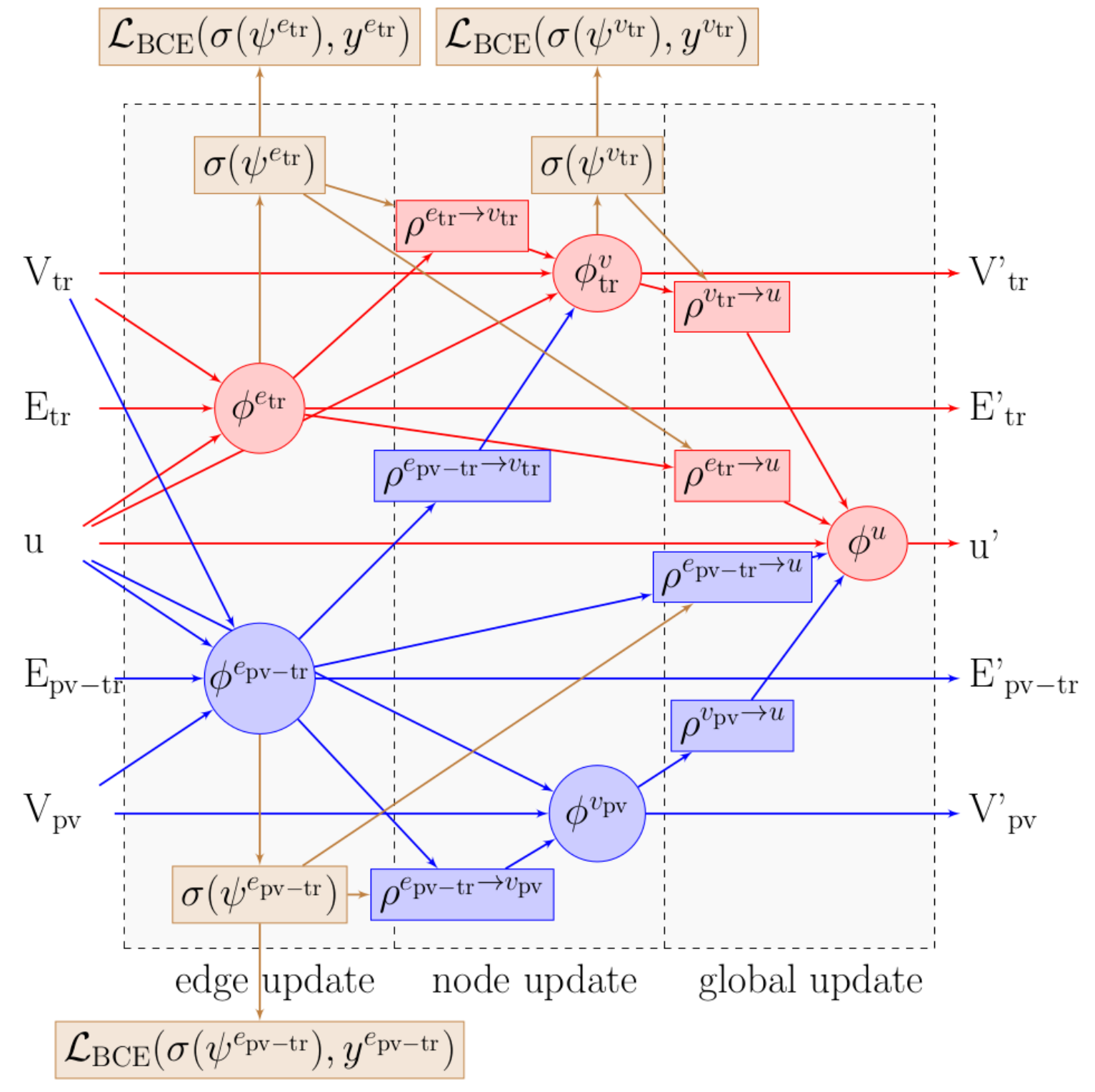}\put(-240,200){(b)}
\end{center}
\caption{(a) Heterogeneous graph representation (above) and HGNN architecture (below) for a simultaneous beauty hadron reconstruction and PV association. In the graph, solid lines (true edges) are shown with their class labels and indicate a physical relationship between nodes: for PV-track edges, the track is associated with the corresponding PV, while for track-track edges, the tracks originate from the same beauty-hadron decay. Dashed lines (false edges) denote examples of connections where no such relationship exists and have a label 0. (b) Heterogeneous modification in blue to the GNN layer updates from Battaglia {\it et al}.~\cite{battaglia2018relationalinductivebiasesdeep} in red. Here, the inputs to the HGNN layer are the sets of node features for tracks and PVs, $V_{\rm tr}$ and $V_{\rm pv}$, sets of edge features for track-track and PV-track edges, $E_{\rm tr}$ and $E_{\rm pv-tr}$, and finally, global features, $u$. The sequence of computations corresponding to Equation~\ref{eq:GNeq2} is illustrated, while the added pruning tasks are indicated in light brown.}
\label{fig:hetero}
\vskip -0.15in
\end{figure}

Figure~\ref{fig:hetero}(a) illustrates the heterogeneous nature of a LHCb collision event with multiple reconstructed PVs, which produce $\mathcal{O}$(100-1,000) charged particle tracks. Data relations within the collision event can be naturally represented by a heterogeneous graph, with PV and track node types, and track-track and PV-track edges, which signify relations between various node objects. Here, we chose not to include PV-PV edges, as they are not directly associated with any learning objective. Our previous DFEI algorithm assigned each track to a PV with a minimum impact parameter and appended the corresponding PV coordinates as static track‑node features. This capped performance, because the model could not learn to select the correct PV dynamically, and increased memory requirements by duplicating the same coordinates across tracks. By contrast, a heterogeneous graph explicitly encodes PV-track edges, allowing end‑to‑end learning of associations and eliminating repetitions of PV data.

To enable learning of graph representations for joint beauty-hadron reconstruction and PV association, we propose a heterogeneous extension of the GNN layer of Battaglia {\it et al.}~\cite{battaglia2018relationalinductivebiasesdeep}. We denote the track and PV node representations, $v^{i}_{\rm tr}$ and $v^{j}_{\rm pv}$, and the track-track and PV-track edge representations, $e^{k}_{\rm tr}$ and $e^{l}_{\rm pv-tr}$. The corresponding edge, node and global representation updates for the HGNN layer are as follows:
\begin{equation}
\begin{split}
e'^{k}_{\rm tr} = \phi^{e_{\rm tr}}(e^{k}_{\rm tr}, v^{r_{k}}_{\rm tr}, v^{s_{k}}_{\rm tr}, u) \qquad & \bar{e}'^{i}_{\rm tr} = \rho^{e_{\rm tr}\rightarrow v_{\rm tr}}(\{ E'^{i}_{\rm tr} \})\\
e'^{l}_{\rm pv-tr} = \phi^{e_{\rm pv-tr}}(e^{l}_{\rm pv-tr}, v^{m_{l}}_{\rm tr}, v^{n_{l}}_{\rm pv}, u) \qquad  & \bar{e}'^{i}_{\rm pv-tr} = \rho^{e_{\rm pv-tr}\rightarrow v_{\rm tr}}(\{ E'^{i}_{\rm pv-tr} \})\\
v'^{i}_{\rm tr} = \phi^{v_{\rm tr}}(v^{i}_{\rm tr}, \bar{e}'^{i}_{\rm tr},\bar{e}'^{i}_{\rm pv-tr} ,   u) \qquad &  \bar{e}'^{j}_{\rm pv-tr}=\rho^{e_{\rm pv-tr}\rightarrow v_{\rm pv}}(\{ E'^{j}_{\rm pv-tr} \})\\
v'^{j}_{\rm pv} = \phi^{v_{\rm pv}}(v^{j}_{\rm pv},\bar{e}'^{j}_{\rm pv-tr},  u) \qquad &  \bar{e}'_{\rm tr} = \rho^{e_{\rm tr}\rightarrow u}(\{ E'_{\rm tr}  \}), \bar{v}'_{\rm tr} = \rho^{v_{\rm tr}\rightarrow u}(\{ V'_{\rm tr} \})\\
u' = \phi^u(\bar{e}'_{\rm tr}, \bar{e}'_{\rm pv-tr}, \bar{v}'_{\rm tr}, \bar{v}'_{\rm pv}, u)  \qquad & \bar{e}'_{\rm pv-tr} = \rho^{e_{\rm pv-tr}\rightarrow u}(\{ E'_{\rm pv-tr}  \}), \bar{v}'_{\rm pv} = \rho^{v_{\rm pv}\rightarrow u}(\{ V'_{\rm pv} \})\; ,
\end{split}
\label{eq:GNeq2}
\end{equation}
where the HGNN layer update begins with edge updates $e'^{k}_{\rm tr}$ and $e'^{l}_{\rm pv-tr}$, applying MLPs $\phi^{e_{\rm tr}}$ and $\phi^{e_{\rm pv-tr}}$. Meanwhile, node updates $v'^{i}_{\rm tr}$ and  $v'^{j}_{\rm pv}$ with MLPs $\phi^{v_{\rm tr}}$ and $\phi^{v_{\rm pv}}$ subsequently take aggregated edge representations for multiple edge types as input. The inputs of the $v'^{i}_{\rm tr}$ update include the aggregated track edges $\bar{e}'^{i}_{\rm tr}$ and PV-track edges $\bar{e}'^{i}_{\rm pv-tr}$. Meanwhile, the update $v'^{j}_{\rm pv}$ includes the aggregation of PV-track edges to PV nodes, $\bar{e}'^{j}_{\rm pv-tr}$. Finally, the global update uses aggregations of all node and edge types as input, which include $\bar{e}'_{\rm tr}$, $\bar{e}'_{\rm pv-tr}$, $\bar{v}'_{\rm tr}$, $\bar{v}'_{\rm pv}$. Figure~\ref{fig:hetero}(b) shows the various HGNN layer computations of Equation~\ref{eq:GNeq2} for node and edge inputs from tracks (red) and PVs (blue).

To enhance the scalability of the HGNN layer in potential real-time data acquisition settings, we incorporate edge and node pruning directly into its layers. The goal is to learn node and edge scores that can be used to remove irrelevant nodes and edges at inference time. To this end, we introduce node and edge scores, $\hat{y}^{v} = \sigma(\psi^{v}(v'))$ and $\hat{y}^{e} = \sigma(\psi^{e}(e'))$, where $\psi^{v}$ and $\psi^{e}$ are MLPs operating on the updated node and edge representations $v'$ and $e'$ from Equation~\ref{eq:GNeq2}, and $\sigma$ denotes the sigmoid function. These probability scores can be trained in two ways: using a Binary Cross-Entropy loss $\mathcal{L}_{\rm BCE}(\hat{y}^{v/e}, y^{v/e})$ when domain knowledge provides suitable ground-truth pruning labels $y^v$ and $y^e$, or implicitly, by using the predicted scores $\hat{y}^{v}$ and $\hat{y}^{e}$ as weights within the HGNN aggregation functions. We explore both approaches in our ablation studies in Section~\ref{sec:ablation}.

Pruning can be performed either via a top-$k$ operation or by introducing a threshold $y_{\rm cut}$. In this paper, we adopt the latter approach, given its better observed performance and greater interpretability as the threshold $y_{\rm cut}$ directly defines the purity of the retained subgraph, i.e. the fraction of nodes and edges considered relevant according to their predicted scores. During training, these scores are used as continuous weights within the HGNN aggregation functions rather than for hard removal, effectively encouraging the model to favor sparse and informative connections while retaining differentiability. The resulting sparsity bias leads to improved reconstruction performance by suppressing redundant background contributions. At inference time, a hard pruning with $y_{\rm cut}$ is applied to remove redundant background edges and nodes while further increasing processing speed and reduce memory usage.  We acknowledge that integrating pruning into training (i.e. pruning-aware or dynamic graph construction) could further improve reconstruction quality. Nonetheless, our empirical results show that incorporating additional MLPs per node and edge already improves the algorithm’s performance, as detailed in the subsequent sections.

\subsection{HGNN architecture and objective loss}
\label{sec:architecture}

The full HGNN architecture is shown in Figure~\ref{fig:hetero}(a), which processes the input $G_{\rm in}$ including track and PV node and edge input features through an encoder layer, several HGNN layers with integrated graph pruning and a decoder layer resulting in $G_{\rm out}$. The encoder and decoder layers consist of MLP node, edge and global updates $\chi(v)$, $\chi(e)$ and $\chi(u)$. The encoded graph representation, $G_{\rm enc}$, is concatenated with the output graph representation from each HGNN layer to preserve important information from the input features and prevent oversmoothing~\cite{Li:2018deeper}. The HGNN was implemented as a custom architecture using the PyTorch framework~\cite{pytorch}, and the code for training and performance studies is publicly available~\cite{github}.

The hyperparameters of the HGNN include the number of HGNN layers ($n^{\rm HGNN}_{\rm layers} = 8$) and layers within MLPs ($n^{\rm MLP}_{\rm layers} = 4$), MLP hidden channel dimensions ($d^{\chi}_{\rm h} = d^{\phi}_{\rm h} = 128$ and $d^{\psi}_{\rm h} = 16$) and MLP output dimensions ($d^{\chi}_{\rm h} = d^{\phi}_{\rm h} = 16$ and $d^{\psi}_{\rm h} = 1$). All MLPs employ rectified linear unit (ReLU) activation functions and batch norms, whereas aggregation functions use summation. The hyperparameters were selected based on the observed saturation in the validation loss with model complexity.

The HGNN is trained using stochastic gradient descent with the Adam optimisation routine~\cite{kingma2015adam} to accomplish beauty-hadron reconstruction, graph pruning and PV association tasks by minimising a multi-objective loss,
\begin{equation}
\begin{split}
\mathcal{L} =& \mathcal{L}_{\rm CE}\left(e_{\rm out}, y^{\rm LCA}\right) + \beta^{e_{\rm tr}} \sum^{n^{\rm GNN}_{\rm layers}}_{i} \mathcal{L}_{\rm BCE}\left(\hat{y}^{e_{\rm tr}}_{i}, y^{e_{\rm tr}}_{i}\right) +\beta^{v_{\rm tr}} \sum^{n^{\rm GNN}_{\rm layers}}_{i} \mathcal{L}_{\rm BCE}\left(\hat{y}^{v_{\rm tr}}_{i}, y^{v_{\rm tr}}_{i}\right) \\
&+  \beta^{e_{\rm pv-tr}} \sum^{n^{\rm GNN}_{\rm layers}}_{i} \mathcal{L}_{\rm BCE}\left(\hat{y}^{e_{\rm pv-tr}}_{i}, y^{e_{\rm pv-tr}}_{i}\right)
\end{split}
\label{eq:loss}
\end{equation}
where in the first term $\mathcal{L}_{\rm CE}$ denotes a class weighted cross-entropy loss, $e_{\rm out}$ is the output HGNN edge representation and $y^{\rm LCA}$ is a multi-class target for the LCA reconstruction task. The subsequent terms represent the edge and node pruning tasks introduced in Section~\ref{sec:HGNN}, where the significance of the corresponding pruning task relative to the LCA task is scaled with $\beta$ parameters. The pv-tr edge pruning task effectively performs a PV association  by selecting the pv-tr edge with the highest probability score $\hat{y}^{e_{\rm pv-tr}}$ for a given track. By default trainings used $\beta^{e_{\rm tr}}=33$, $\beta^{v_{\rm tr}}=1$ and $\beta^{e_{\rm pv-tr}} =3$. Although a comprehensive optimisation of these hyperparameters was beyond the scope of the study, the sensitivity of the multi-objective optimisation to $\beta^{e_{\rm tr}}$ and $\beta^{v_{\rm tr}}$ is further investigated in ~\ref{sec:hyper}. 

We benchmark the HGNN architecture against a GNN using GNN layer updates according to Equation~\ref{eq:GNeq1} and an equivalent pruning mechanism. The GNN only considers tracks  and their interconnections as nodes and edges. Meanwhile, it uses the same objective loss and hyperparameter values but with the removal of the pv-tr edge pruning task. For both the GNN and HGNN architectures, we also quantify the inclusion of weighted aggregations during message passing and denote the corresponding architectures as WGNN and WHGNN.

\subsection{Datasets and training procedure}

\label{sec:training}

The model is trained, validated and tested using the publicly available datasets from our previous publication DFEI~\cite{dfei_dataset}. In addition, we provide an updated format of the relevant datasets~\cite{dataset}. The datasets were produced with a custom simulation environment using PYTHIA8~\cite{Bierlich:2022pfr} and EvtGen~\cite{Ryd:2005zz} to simulate the particle collision conditions anticipated for LHCb Run 3. Only events containing at least one beauty hadron are considered; these hadrons then decay with EvtGen into a variety of known modes. All results in this paper focus exclusively on charged particles produced within the LHCb geometrical acceptance and Vertex Locator region. Particles outside these regions and neutral particles were not considered, meaning that they are also excluded from the ground truth heavy-hadron decay chains. Additional exclusive datasets, in which one of the beauty hadrons is required to decay to several exclusive decays, are used to further evaluate the model performance.

To emulate the response of the LHCb detector, the generated vertices and particles are further processed to reflect reconstruction effects. Primary vertices with fewer than four associated charged particles are discarded, while the remaining vertex positions are smeared according to LHCb Run 2 resolutions from Reference~\cite{LHCb:2018zdd}. Each particle’s origin point, corresponding to the first hit in the Vertex Locator, is projected onto the nearest of 52 planes representing the detector geometry, with additional Gaussian smearing of 8.5 $\mu$m applied in the $x$ and $y$ directions. Particle momentum directions were smeared according to the momentum-dependent angular resolutions from Reference~\cite{Billoir:2021srr}, while the momentum magnitude was smeared with a relative resolution of 0.4\%~\cite{LHCb:2014uqj}. This procedure captures the main features of LHCb reconstruction, although secondary particles from material interactions and misreconstructed tracks are not included.

As depicted in Figure~\ref{fig:hetero}, collision events are represented as a heterogeneous graph. Each node and edge type has an input-feature representation. For charged particles (tracks), the input features include the track origin point, $(x_{\rm tr}, y_{\rm tr}, z_{\rm tr})$, track momenta ($p_{x}, p_{y}, p_{z}$) and track charge $q$. Meanwhile, PV nodes are represented by their position coordinates, $(x_{\rm pv}, y_{\rm pv}, z_{\rm pv})$. In contrast, for the GNN benchmarks, we adopted DFEI's approach of appending the coordinates of the PV with the minimum impact parameter to the features of each track. Although highly discriminating timing information is anticipated for PV association in LHCb Upgrade II~\cite{LHCb:2018roe}, this study limited its scope to using only the previously defined positional and kinematic features.

A fully connected graph is defined between all track and PV nodes with track-track and PV-track edges. The input edge representation for PV-track edges includes only the impact parameter of the track with respect to the PV. For track-track edge features include the angle $\theta$ between the three-momentum directions of the two particles, the difference $\Delta z_{\rm tr}$ in origin $z_{\rm tr}$, a Boolean indicating a shared PV according to the minimum impact parameter, and the momentum-transverse distance, which is the distance between the origin points in a plane transverse to the momentum direction. We adopted a loose prefiltering of track-track edges with 99\% efficiency from the DFEI~\cite{GarciaPardinas:2023pmx}, which requires edges to satisfy $\theta < 0.26$ rad. or the condition of a shared PV according to the minimum impact parameter. For the remaining edges, the ground-truth LCAG is determined $y^{\rm LCA}$, where $y^{\rm LCA}=0$ indicates no shared ancestors. The non-zero values were restricted to $y^{\rm LCA}=1,2,3$, which were found to be sufficient to represent the decay hierarchies present in the simulation dataset. Finally, global input features include the multiplicity of tracks and PVs in the event.

The training and validation datasets consisted of 40,000 and 10,000 events respectively, which are equivalent to those used in our previous work. Nominal training for the ablation studies was performed on an NVIDIA L40S GPU with a batch size of 12. For ablation studies, the base schedule consisted of 30 epochs at a learning rate of $\alpha = 10^{-3}$, followed by two epochs at $\alpha = 10^{-4}$. Before evaluating the reconstruction performance of the selected architectures, we performed an additional fine-tuning phase on the same dataset, consisting of two epochs with a learning rate of $\alpha = 5\times10^{-5}$ followed by two epochs with $\alpha = 2\times10^{-5}$. During all training phases, training and validation losses as well as classification accuracies were monitored (see ~\ref{sec:appendixA} for examples of training and validation loss curves). The final performance was measured using a test set of 10,000 independent inclusive beauty-hadron decay events and multiple exclusive decay datasets, each containing 5,000 events.

%% file: results.tex
\subsection{Ablation studies}
\label{sec:ablation}
To quantify the relative merits of the novel architectural developments, several ablation models were trained under similar conditions, as described in Section~\ref{sec:training}. The model architectures are either GNN or HGNN, and W signifies the use of the weighted message passing introduced in Section~\ref{sec:HGNN}. Ablations further quantify the addition or removal of track node and edge pruning ($\mathcal{L}^{\rm prune}_{\rm BCE}$) and PV association ($\mathcal{L}^{\rm PV}_{\rm BCE}$) tasks by removing the corresponding BCE loss terms from Equation~\ref{eq:loss} during training.

\begin{table}[!h]
\begin{tabular}{llrllll}

Model & Tasks & $\mathcal{L}^{\rm LCA}_{\rm CE}$ & $y^{\rm LCA}=0$ & $y^{LCA}=1$ & $y^{LCA}=2$ & $y^{LCA}=3$ \\
\br
GNN & $\mathcal{L}^{\rm LCA}_{\rm CE}$ & 0.56 & 98.220 ± 0.001 & 68.0 ± 0.2 & 55.2 ± 0.1 & 79.9 ± 0.2 \\
GNN & $\mathcal{L}^{\rm LCA}_{\rm CE}, \mathcal{L}^{\rm prune}_{\rm BCE}$ & 0.49 & 99.441 ± 0.001 & 75.5 ± 0.2 & 60.3 ± 0.1 & 83.2 ± 0.2 \\
WGNN & $\mathcal{L}^{\rm LCA}_{\rm CE}$ & 0.60 & 97.955 ± 0.001 & 63.2 ± 0.2 & 53.5 ± 0.1 & 76.1 ± 0.3 \\
WGNN & $\mathcal{L}^{\rm LCA}_{\rm CE}, \mathcal{L}^{\rm prune}_{\rm BCE}$ & 0.47 & 99.282 ± 0.001 & 76.9 ± 0.2 & 57.9 ± 0.1 & 85.6 ± 0.2 \\
HGNN & $\mathcal{L}^{\rm LCA}_{\rm CE}$ & 0.54 & 98.826 ± 0.001 & 71.3 ± 0.2 & 51.6 ± 0.1 & 80.9 ± 0.2 \\
HGNN & $\mathcal{L}^{\rm LCA}_{\rm CE},\mathcal{L}^{\rm PV}_{\rm BCE}$ & 0.53 & 98.870 ± 0.001 & 71.8 ± 0.2 & 52.7 ± 0.1 & 82.5 ± 0.2 \\
HGNN & $\mathcal{L}^{\rm LCA}_{\rm CE},\mathcal{L}^{\rm PV}_{\rm BCE}, \mathcal{L}^{\rm prune}_{\rm BCE}$ & 0.49 & 99.289 ± 0.001 & 75.8 ± 0.2 & 61.4 ± 0.1 & 83.9 ± 0.2 \\
WHGNN & $\mathcal{L}^{\rm LCA}_{\rm CE}$ & 0.58 & 98.683 ± 0.001 & 68.5 ± 0.2 & 52.8 ± 0.1 & 76.7 ± 0.2 \\
WHGNN & $\mathcal{L}^{\rm LCA}_{\rm CE},\mathcal{L}^{\rm PV}_{\rm BCE}$ & 0.51 & 98.959 ± 0.001 & 71.7 ± 0.2 & 54.8 ± 0.1 & 83.2 ± 0.2 \\
WHGNN & $\mathcal{L}^{\rm LCA}_{\rm CE},\mathcal{L}^{\rm PV}_{\rm BCE}, \mathcal{L}^{\rm prune}_{\rm BCE}$ & 0.46 & 99.274 ± 0.001 & 75.9 ± 0.2 & 61.3 ± 0.1 & 84.0 ± 0.2 \\
\br
\end{tabular}
\caption{Comparison of the LCAG loss value and class accuracies in percent on the test dataset for various architectural ablations. The uncertainties on the LCAG class accuracies are statistical in nature.}
\label{table:ablation}
\end{table}

Table~\ref{table:ablation} shows the various performance metrics, including the LCAG loss $\mathcal{L}^{\rm LCA}_{\rm CE}$ and class accuracies, as evaluated with the test dataset. Incorporating additional tasks consistently improves the accuracy of LCAG reconstruction, with the largest gains seen when including track edge and node pruning tasks. Although the GNN and HGNN architectures achieve comparable LCAG reconstruction performance, only the HGNN supports PV association, the results of which are presented in Section \ref{sec:pvassociation}. Based on the ablation studies we only present subsequent performance results for the GNN and HGNN architectures trained with all possible tasks.

Uncertainties on the LCAG class accuracies are statistical and correspond to standard deviations estimated under a binomial model. 
For each class $y^{\rm LCA}=i$, the accuracy is computed as the fraction of correctly classified examples 
$N^{\text{correct}}_{y^{\rm LCA}=i} / N_{y^{\rm LCA}=i}$, and the uncertainty is given by 
$\sigma = \sqrt{p(1 - p) / N_{y^{\rm LCA}=i}}$ with $p = N^{\text{correct}}_{y^{\rm LCA}=i} / N_{y^{\rm LCA}=i}$.


\begin{figure}[!h]
\begin{center}
\includegraphics[width=8.2cm]{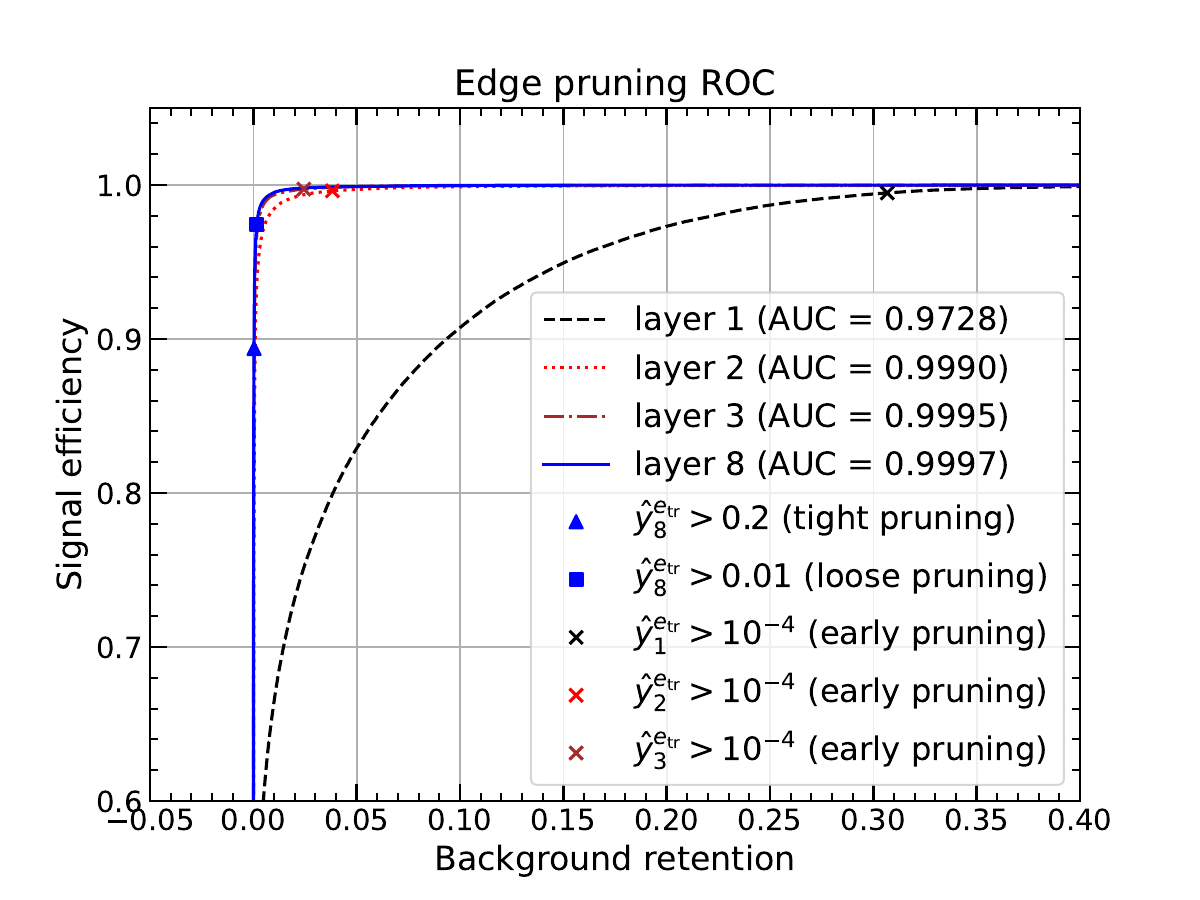}\put(-230,150){\footnotesize(a)}\includegraphics[width=8.2cm]{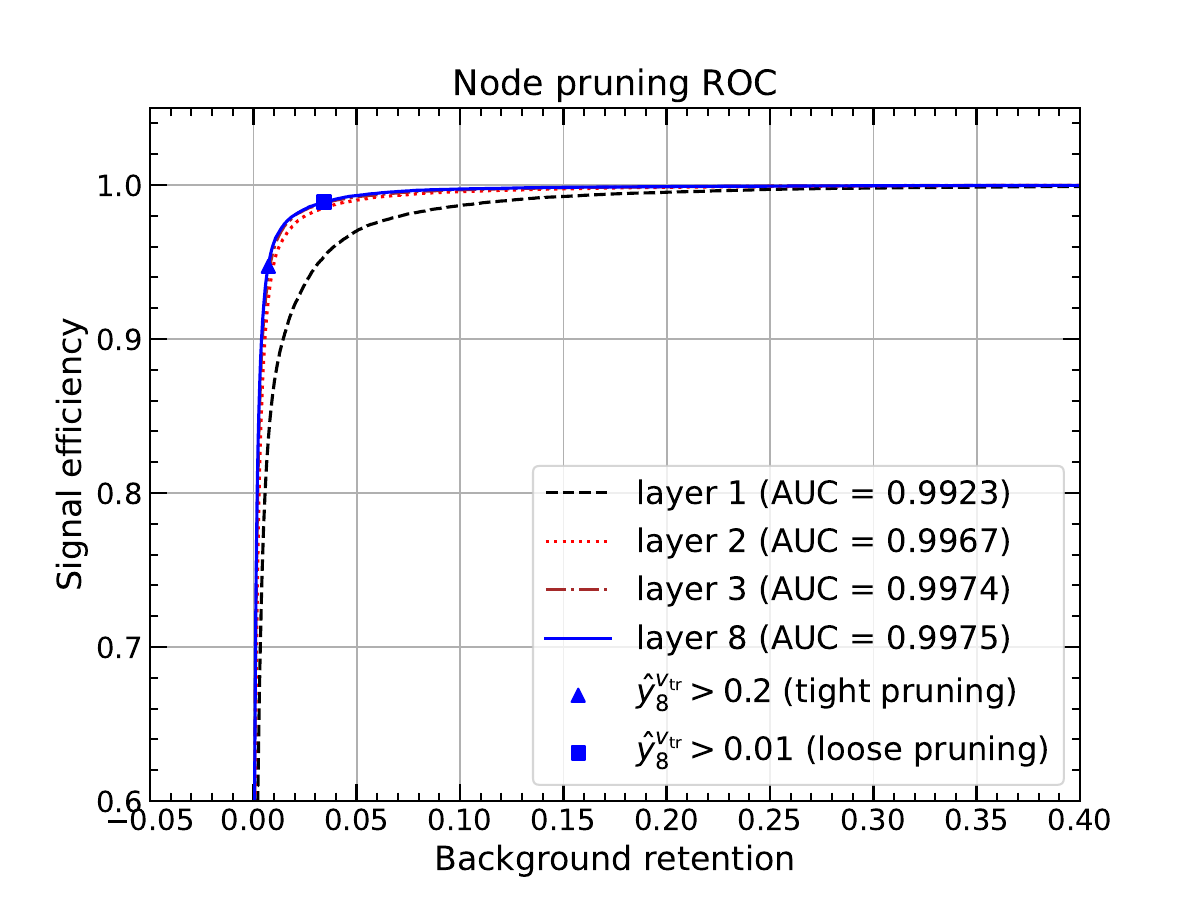}\put(-230,150){\footnotesize(b)} \\
\includegraphics[width=7.8cm]{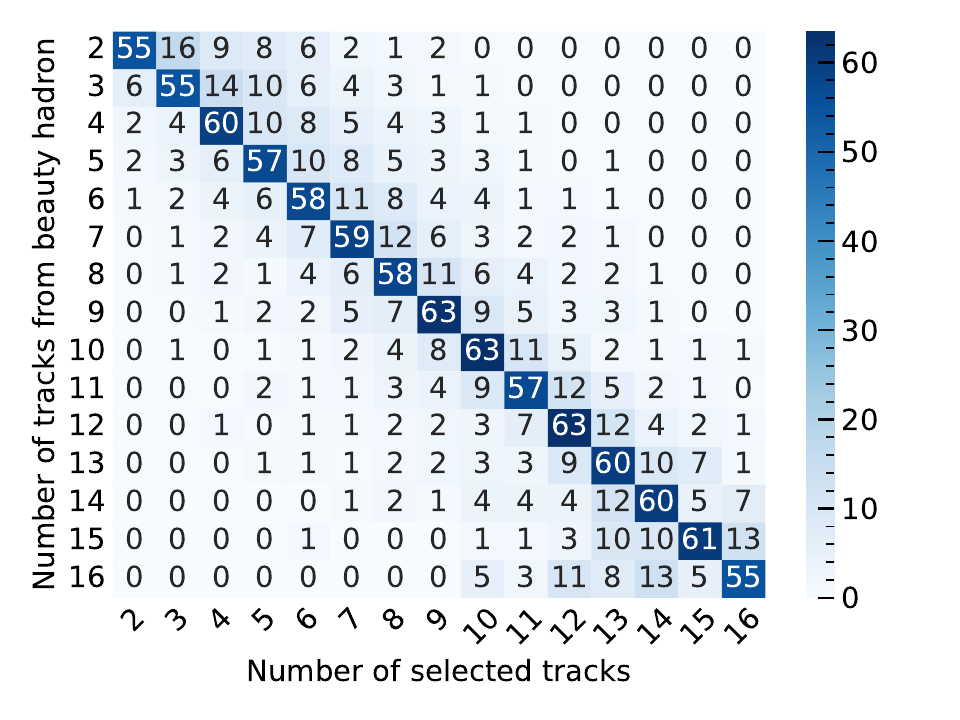}\put(-230,150){\footnotesize(c)}\includegraphics[width=8cm]{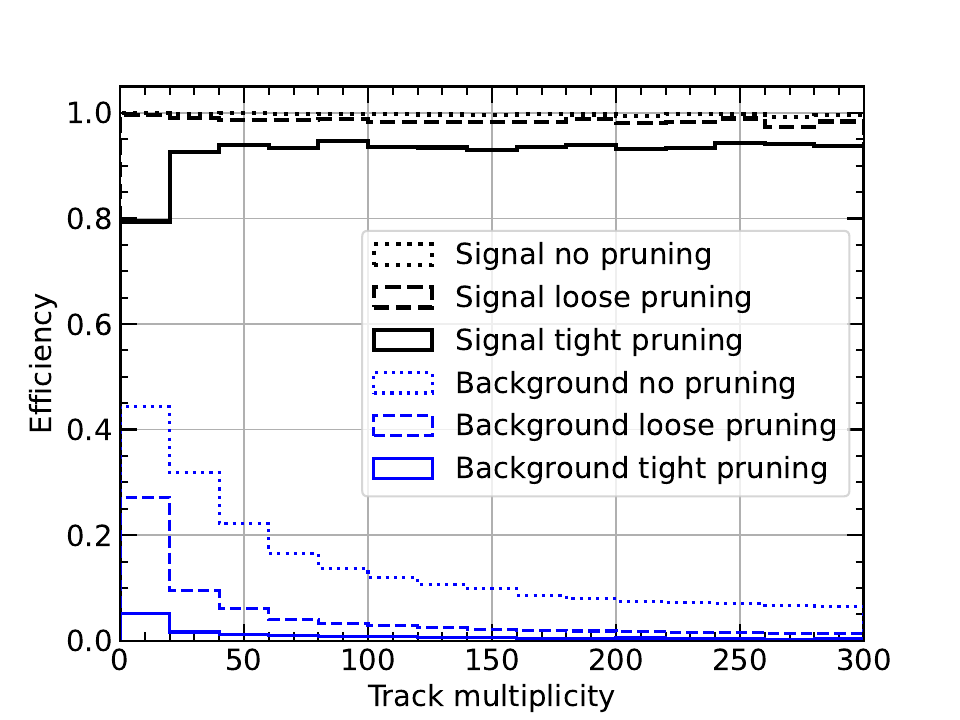}\put(-230,150){\footnotesize(d)}
\end{center}
\caption{ (a) and (b) Receiver Operating Characteristic curves for edge and node scores $\hat{y}^{e_{\rm tr}}$ and $\hat{y}^{v_{\rm tr}}$ at layers 1, 2, 3 and 8. The pruning selections used in the subsequent results are highlighted. (c) Confusion matrix (in percent) between the number of particles originating from beauty hadrons and the number of particles selected with a tight edge and node pruning, $\hat{y}^{e_{\rm tr}}_{8}>0.2$ and $\hat{y}^{v_{\rm tr}}_{8}>0.2$. (d) Track signal efficiency and background retention as functions of track multiplicity for tight and loose ($\hat{y}^{e_{\rm tr}}_{8}>0.01$ and $\hat{y}^{v_{\rm tr}}_{8}>0.01$) pruning selections.}
\label{fig:ROC}
\end{figure}

\subsection{Pruning performance}
\label{sec:pruning}

Figure~\ref{fig:ROC}(a) and (b) show the discrimination power of the track edge and node scores $\hat{y}^{e_{\rm tr}}$ and $\hat{y}^{v_{\rm tr}}$ at HGNN layers 1, 2, 3 and 8. Here, signal edges or nodes (tracks) refer to those originating from a beauty-hadron decay, whereas background edges and nodes correspond to those not associated with such decays. The discrimination power increases significantly between layers 1 and 2 particularly for the edges, with the area under the curve (AUC) increasing from 0.973 to 0.999. While the discrimination power saturates with subsequent layers, there are marginal increases in the discrimination power for each subsequent HGNN layer.

Figure~\ref{fig:ROC}(c) and (d) further quantify the performance of the pruning procedure after applying a selection requiring the LCAG class prediction to be above zero. In particular, Figure~\ref{fig:ROC}(c) effectively serves as a confusion matrix, expressed in percent, illustrating how the number of true tracks from beauty hadrons compares with the number of tracks selected under the tight pruning requirement, $\hat{y}^{e_{\rm tr}}_{8} >0.2$ and $\hat{y}^{v_{\rm tr}}_{8} >0.2$. The diagonal dominance demonstrates that the pruning selection preserves the correct track multiplicity for most beauty decays while effectively suppressing spurious tracks. Small off-diagonal entries reflect cases where either additional non-beauty tracks are incorrectly selected or some beauty tracks are missed. Meanwhile, Figure~\ref{fig:ROC}(d) compares the efficiencies for signal and background tracks as a function of track multiplicity, using the tight pruning selection and a looser pruning selection with $\hat{y}^{e_{\rm tr}}_{8}>0.01$ and $\hat{y}^{v_{\rm tr}}_{8}>0.01$. The efficiencies are defined as the ratio of surviving signal or background tracks after pruning to those before the pruning selection. Tight pruning strongly suppresses background tracks, particularly at high multiplicities, while retaining a large and stable signal efficiency with increasing event complexity. The looser selection preserves nearly full signal efficiency but admits substantially more background, illustrating the trade-off between purity and efficiency achieved by the pruning criteria.

\begin{figure}[!h]
\begin{center}
\includegraphics[width=8.3cm]{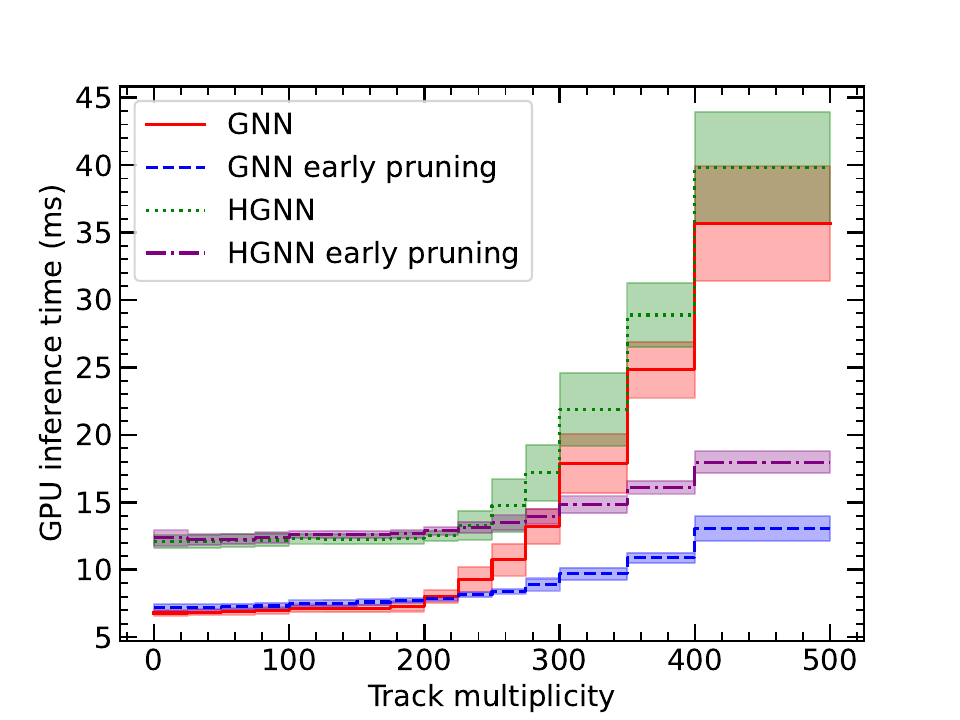}\includegraphics[width=8.3cm]{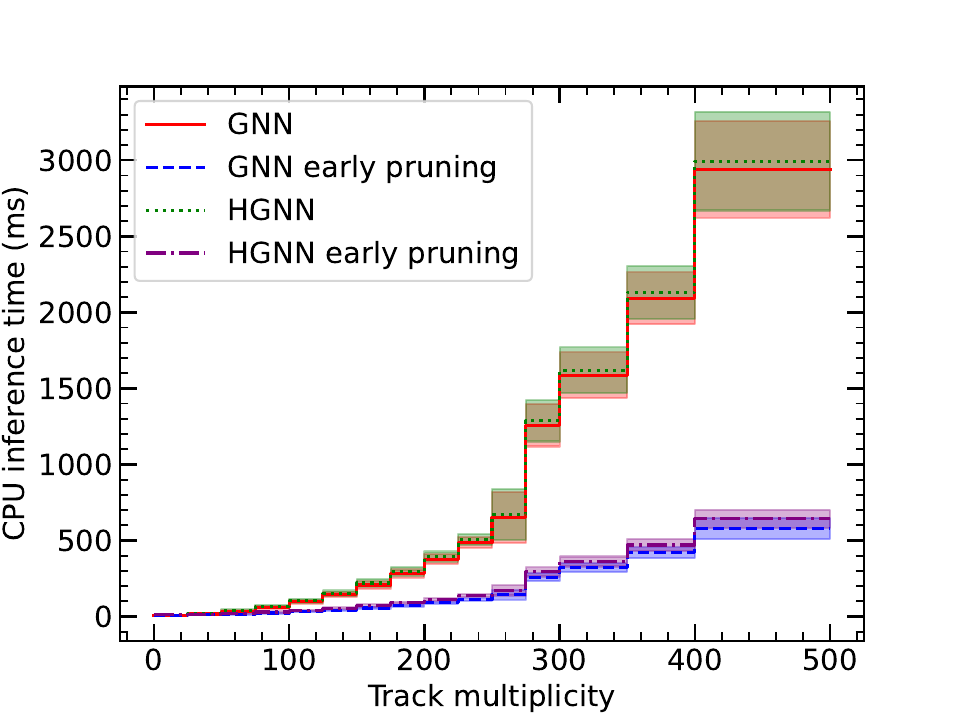}
\caption{GNN and HGNN GPU (left) and CPU (right) mean inference time as a function of track multiplicity per event with and without early pruning ($\hat{y}^{e_{\rm tr} / v_{\rm tr} }_{1-3}>10^{-4}$).The error bands indicate $\pm 1$ standard deviation of the inference time distribution for events in each bin.}
\label{fig:timing}
\end{center}
\end{figure}

The node and edge pruning layers aim to accelerate the inference time of the models, particularly the scaling of inference time with track multiplicity. Figure~\ref{fig:timing} shows GPU and single-threaded CPU inference times, measured with a NVIDIA RTX 4090 GPU and a Intel Core i9-14900K 3.2 GHz CPU, as a function of track multiplicity for the cases of no pruning selection and a loose edge pruning selection at layers 1-3 ($\hat{y}^{e_{\rm tr} / v_{\rm tr} }_{1-3}>10^{-4}$). Early edge pruning yields noticeable speed-ups above approximately 250 tracks; at multiplicities exceeding 400, we observe a 2-3$\times$ acceleration on GPU and a 5$\times$ acceleration on CPU. For example, for events with track multiplicity around 300, single-threaded CPU inference with early pruning takes approximately 300 ms compared to the 2.2 s timing reported for the multi-stage GNN approach in DFEI \cite{GarciaPardinas:2023pmx} on a 2.2 GHz Intel Core processor.

\subsection{Reconstruction performance}

The beauty-hadron reconstruction performance is evaluated by categorising the reconstructed beauty decay chains into several types introduced in ~\cite{GarciaPardinas:2023pmx}: perfect reconstruction (the reconstructed decay products and the decay hierarchy are both correct and there are no missing decay products, as shown in Figure~\ref{fig:hetero}); complete reconstruction (all decay products are correct with none missing; however, the hierarchy is incorrect); not isolated reconstruction (there are one or more decay products, which are not correct); and partial reconstruction (there are missing decay products). For reconstruction performance, we require that the predicted LCAG class is above zero employ the tight pruning selection, $\hat{y}^{e_{\rm tr}}_{8}>0.2$ and $\hat{y}^{v_{\rm tr}}_{8}>0.2$ introduced in Section~\ref{sec:pruning}.

\begin{table}[!h]
    \centering
    \begin{tabular}{lccccc}
        \hline
        Model  & Perfect reco. (\%) & Complete reco. (\%) & Not isolated (\%) & Part. reco. (\%) \\
        \hline
DFEI & 4.7 ± 0.2 & 6.1 ± 0.2 & 76.1 ± 0.4 & 13.1 ± 0.3 \\
GNN & 21.6 ± 0.4 & 20.8 ± 0.4 & 43.8 ± 0.4 & 13.8 ± 0.3 \\
WGNN & 20.9 ± 0.4 & 20.0 ± 0.4 & 44.9 ± 0.4 & 14.2 ± 0.3 \\
HGNN & 22.4 ± 0.4 & 20.1 ± 0.4 & 44.1 ± 0.4 & 13.4 ± 0.3 \\
WHGNN & 21.5 ± 0.4 & 19.3 ± 0.3 & 45.8 ± 0.4 & 13.5 ± 0.3 \\
        \hline
    \end{tabular}
    \caption{Comparison of the percentage of each reconstruction category in 10,000 inclusive beauty-hadron events using the DFEI method and the GNN/HGNN architectures with tight last layer pruning. Errors show the corresponding statistical uncertainty.}
    \label{table:increco}
\end{table}

\begin{figure}[!t]
\begin{center}
\includegraphics[width=8.8cm]{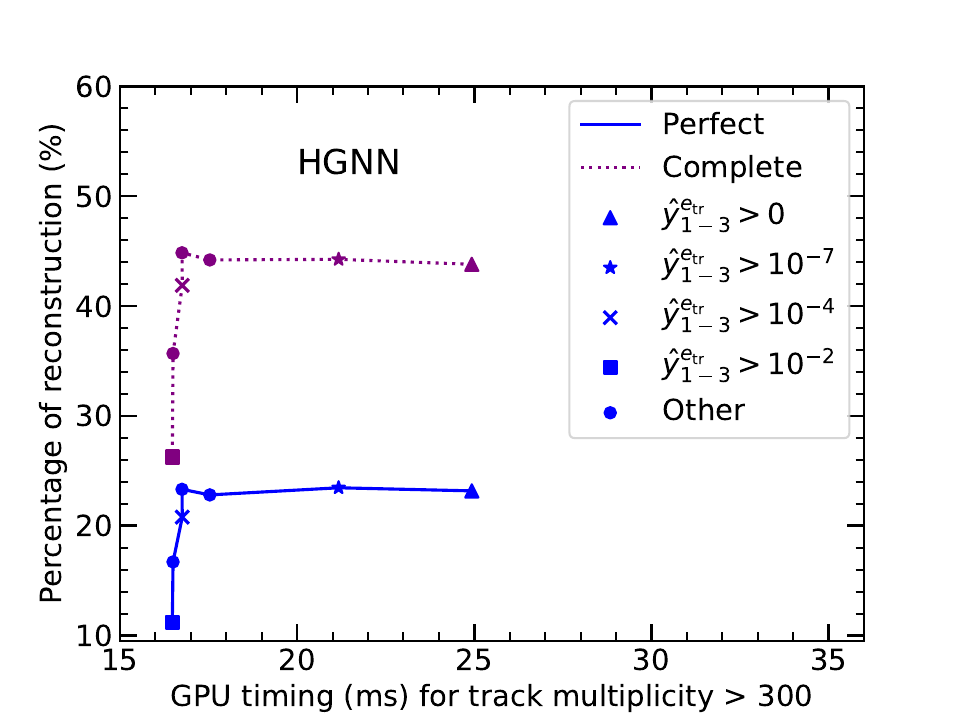}\includegraphics[width=8.8cm]{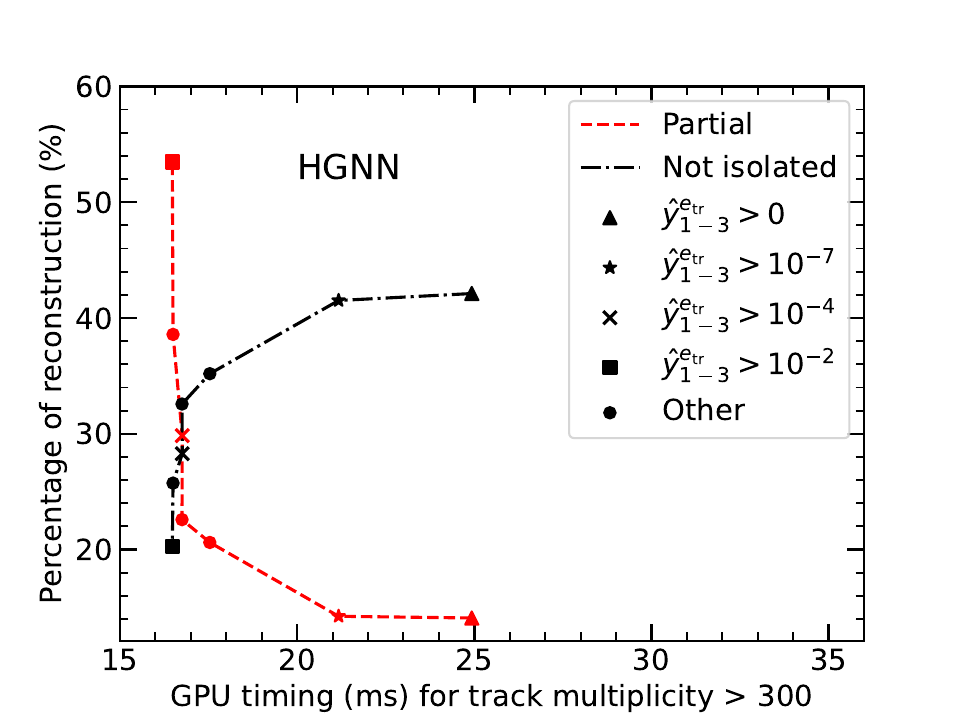} \\
\includegraphics[width=8.8cm]{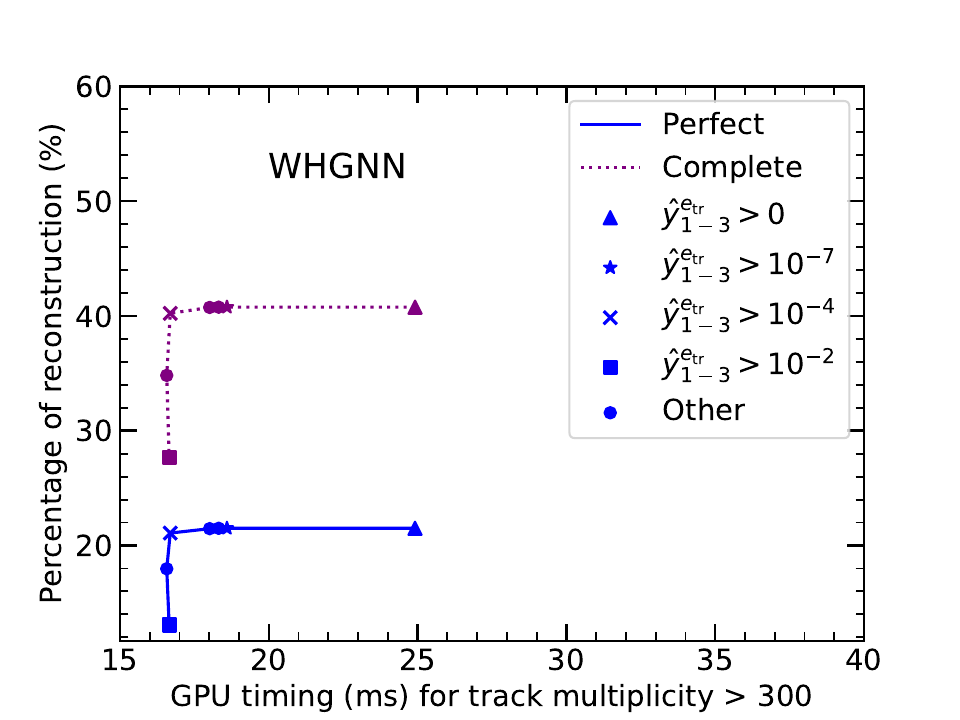}\includegraphics[width=8.8cm]{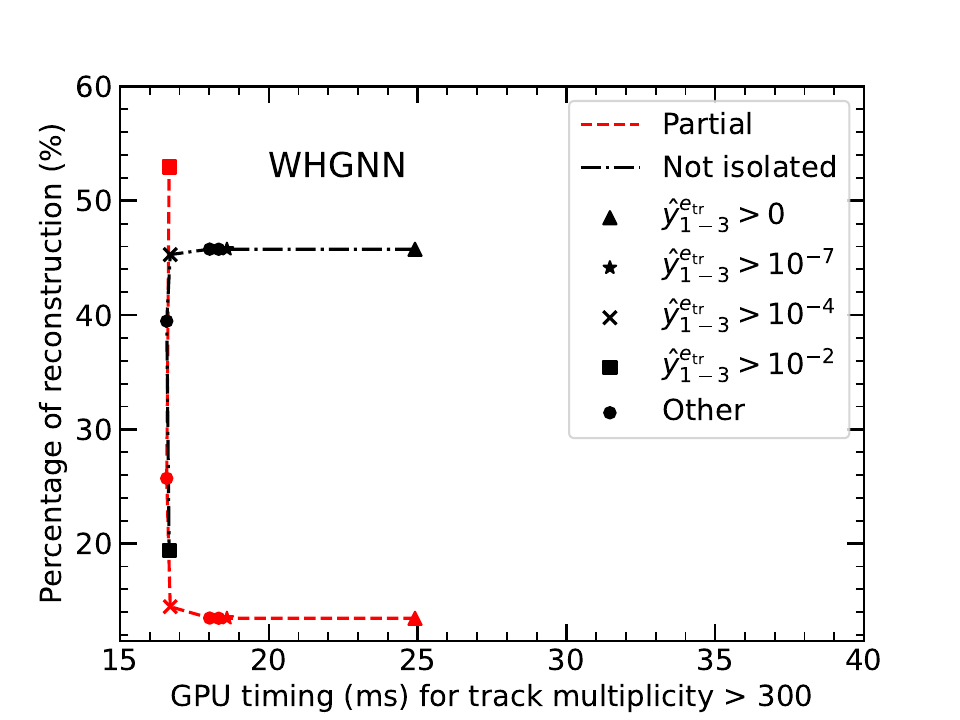}
\caption{Relationship between the average GPU timing (ms) for events with track multiplicity greater than 300 and percentage of each reconstruction category for HGNN (top) and WHGNN (bottom) models. The left panels show the fractions of Perfect and Complete reconstructions, while the right panels show the fractions of Partial and Not isolated reconstructions. The results are shown for configurations without early pruning (corresponding to the triangular points with the highest average GPU timing) and with edge pruning applied in layers 1-3 for selection thresholds ranging from $\hat{y}^{e_{\rm tr}}_{1-3} > 10^{-7}$ to $\hat{y}^{e_{\rm tr}}_{1-3} > 10^{-2}$ in factors of 10.}
\label{fig:timingvsperf}
\end{center}
\end{figure}

Table~\ref{table:increco} compares the reconstruction performance of the DFEI with the proposed GNN and HGNN models for inclusive beauty-hadron collision events. All models significantly outperform the DFEI in terms of the percentage of perfect and complete reconstructions, exhibiting a much lower occurrence of non-isolated decays and a comparable level of partial reconstruction. In particular, the HGNN architecture, which slightly surpasses the other models, achieves a perfect reconstruction rate 4.8 times higher than that of the DFEI. The quoted uncertainties correspond to standard deviations estimated in the same way as for the LCAG class accuracies: for each reconstruction category, the fraction of events in that category is treated as a binomial ratio, giving $\sigma = \sqrt{p(1-p)/N}$ with $p$ the observed fraction and $N$ the total number of events.

When early pruning is used to accelerate the inference time, there can be a significant drop in performance if the pruning cuts are too tight. This is illustrated in Figure~\ref{fig:timingvsperf}, which shows the trade-off between the reconstruction performance and the average GPU inference time for track multiplicities above 300 using the HGNN and WHGNN models. A gradual tightening of pruning selections in layers 1-3 are applied ranging from $\hat{y}^{e_{\rm tr}}_{1-3} > 10^{-7}$ to $\hat{y}^{e_{\rm tr}}_{1-3} > 10^{-2}$ in factors of 10. The WHGNN exhibits greater invariance in performance with pruning, which allows for smaller inference times with minimal reconstruction performance loss. 

\begin{table}[!h]
\centering
\begin{tabular}{c|c|c|c|c|c|c|c|c|c|c|c|c}
\hline
 & \multicolumn{3}{|c|}{Perfect reco.} & \multicolumn{3}{|c|}{Complete reco.} & \multicolumn{3}{|c|}{Not isolated} & \multicolumn{3}{|c}{Part reco.} \\
\hline
Decay & DF & H1 & H2 & DF & H1 & H2 & DF & H1 & H2 & DF & H1 & H2 \\
\hline
inclusive beauty & 4.7 & 22.4 & 21.9 & 6.1 & 20.1 & 20.6 & 76.1 & 44.1 & 44.1 & 13.1 & 13.4 & 13.4 \\
$B^{0}\rightarrow K^{*0} \mu \mu $ & 32.7 & 20.3 & 92.4 & 17.8 & 37.7 & 1.1 & 43.9 & 6.2 & 4.7 & 5.6 & 35.8 & 1.8 \\
$B^{0}\rightarrow K \pi $ & 38.1 & 47.4 & 91.6 & 0.0 & 0.0 & 0.0 & 54.7 & 10.2 & 7.0 & 7.2 & 42.4 & 1.4 \\
$B^{+}\rightarrow K \pi \pi $ & 35.6 & 23.7 & 94.5 & 10.3 & 26.3 & 0.2 & 46.5 & 8.5 & 4.7 & 7.6 & 41.5 & 0.6 \\
$B^{0}_{s} \rightarrow J/\psi \phi$ & 31.3 & 22.8 & 91.8 & 20.3 & 44.3 & 1.7 & 44.3 & 9.9 & 5.0 & 4.1 & 22.9 & 1.5 \\
$\Lambda_{b} \rightarrow \Lambda^{+}_{c}\pi$ & 22.2 & 27.5 & 68.3 & 8.6 & 9.4 & 24.4 & 37.4 & 7.3 & 5.2 & 31.8 & 55.7 & 2.1 \\
\hline
$B^{0} \rightarrow K \mu \mu$ & 36.2 & 21.0 & 93.5 & 10.4 & 28.1 & 0.3 & 45.9 & 8.4 & 4.9 & 7.5 & 42.5 & 1.2 \\
$B^{0}_{s} \rightarrow D^{-}_{s} \pi$ & 33.0 & 57.7 & 67.5 & 7.1 & 11.6 & 23.0 & 53.5 & 13.1 & 7.0 & 6.4 & 17.6 & 2.6 \\
$B^{0} \rightarrow D^{+} D^{-}$ & 26.2 & 37.1 & 56.7 & 23.9 & 40.2 & 32.1 & 45.7 & 14.3 & 7.3 & 4.1 & 8.4 & 4.0 \\
$\Lambda_{b} \rightarrow p K$ & 39.5 & 24.2 & 92.3 & 0.0 & 0.0 & 0.0 & 48.6 & 5.7 & 6.4 & 12.0 & 70.1 & 1.3 \\
$\Lambda_{b} \rightarrow p K \mu \mu$ & 40.9 & 11.5 & 94.7 & 11.1 & 17.7 & 0.5 & 37.4 & 4.8 & 3.7 & 10.6 & 66.1 & 1.1 \\
\hline
\end{tabular}
\caption{Comparison of the percentage of each reconstruction category for DFEI (DF)~\cite{GarciaPardinas:2023pmx} and two HGNN models (H1 and H2) with tight pruning, for several exclusive beauty-hadron decays. Both DFEI and H1 were trained only on inclusive beauty-hadron events, while H2 additionally includes a sample of 3000 events split between the decays listed in the upper half of the table. We consider the following subdecays $D^{+}[K \pi \pi]$, $D^{-}_{s}[K K \pi]$, $J/\psi[\mu \mu]$, $\phi[K K]$, $K^{*0}[K \pi]$ and $\Lambda^{+}_{c}[p K \pi]$.}
\label{table:exclusive}
\end{table}

\begin{figure}[!h]
\begin{center}
\includegraphics[width=10.6cm, trim={0 0 1.8cm 0}, clip]{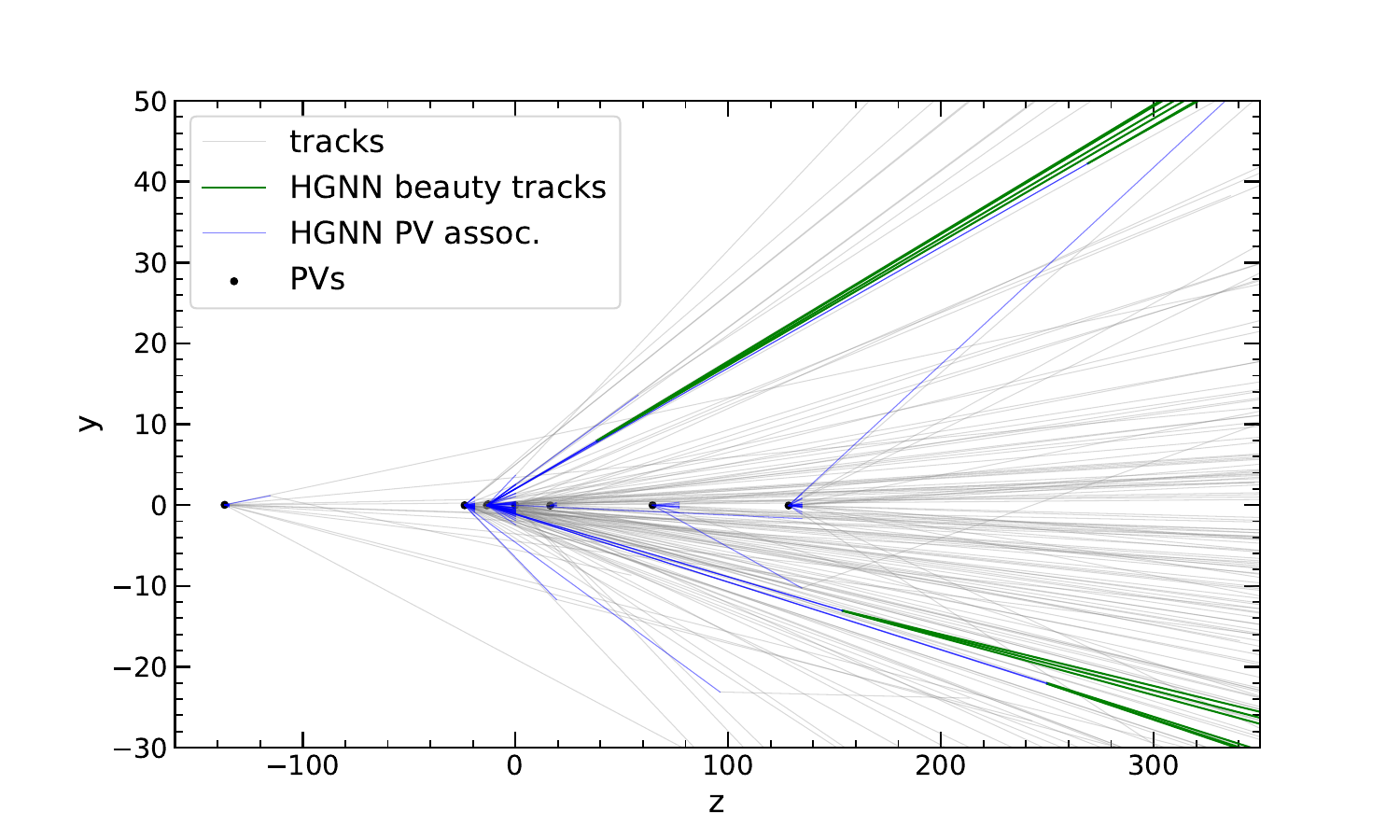}\includegraphics[width=6.2cm]{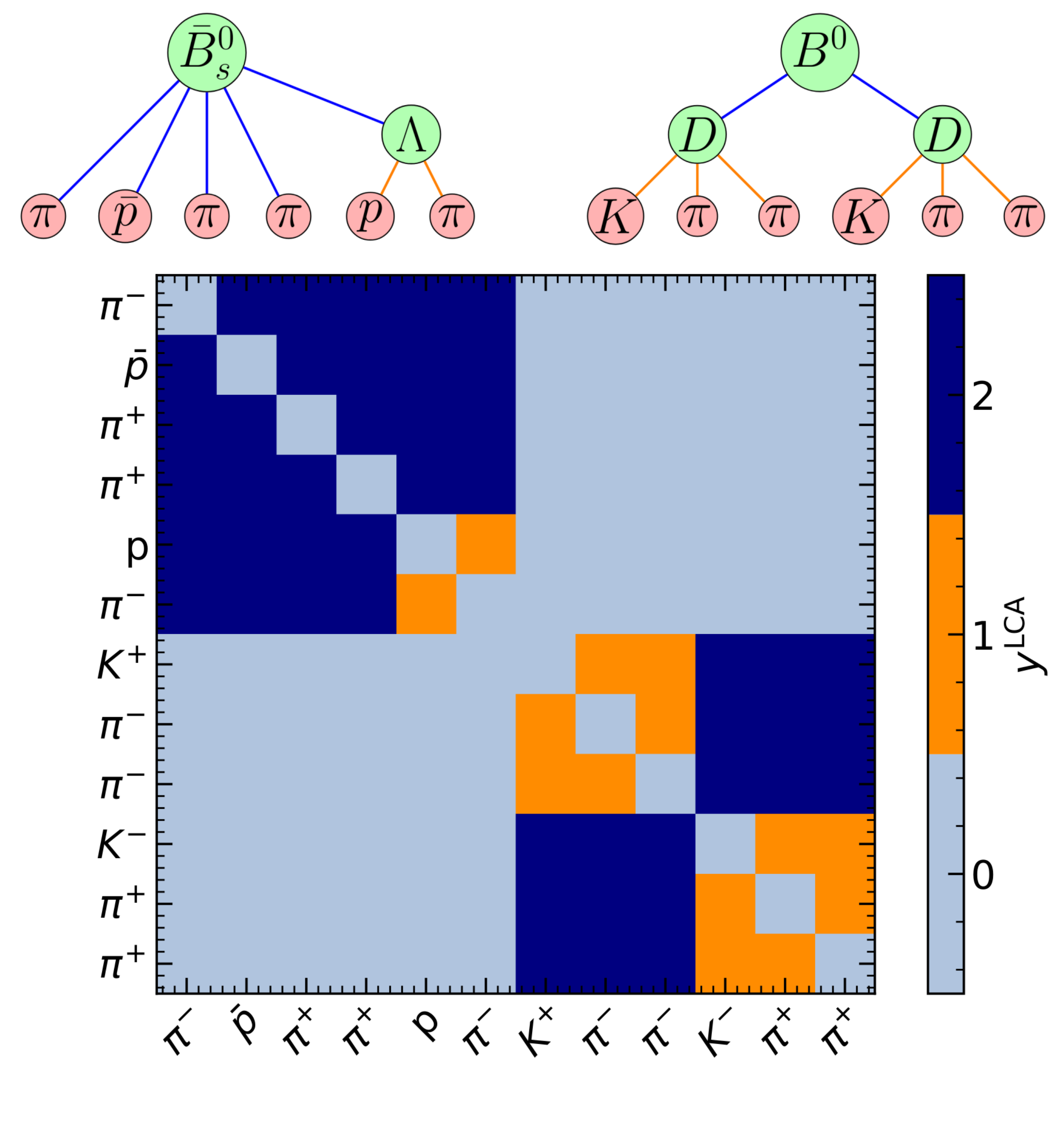}
\end{center}
\caption{Demonstration of the HGNN inference in an event with seven PVs. The HGNN is able to simultaneously associate all tracks to a PV (in blue), isolate the tracks from beauty hadrons via edge and node pruning (in green) and determine the decay hierarchy of both beauty hadrons via its LCAG prediction shown on the right in matrix form. }
\label{fig:event}
\end{figure}

The reconstruction performance can be further quantified for several exclusive beauty-hadron decays with different decay topologies. Table~\ref{table:exclusive} compares the performance of the DFEI model (DF) and HGNN architecture (H1) for a variety of specific decay modes. While the HGNN often outperforms the DFEI in terms of perfect and complete reconstruction categories and has a smaller fraction of not isolated reconstruction, it suffers from larger levels of partial reconstruction. For comparison, an equivalent HGNN (H2) was trained with the addition of 3000 training samples split uniformly between the decay modes, which are shown in the upper half of Table~\ref{table:exclusive}. H2 achieves a much higher percentage of perfect reconstruction, while maintaining a 1-2\% level of partial reconstruction. Moreover, H2 generalises to similar exclusive decays not seen during training, as demonstrated in the lower half of Table~\ref{table:exclusive}.

An example of the HGNN reconstruction is shown in Figure~\ref{fig:event} for an event with seven PVs with one producing $B^{0} \rightarrow D^{+}[K^{+}\pi^{+}\pi^{-}] D^{-}[K^{-}\pi^{+}\pi^{-}]$ and $B^{0}_{s} \rightarrow  \bar{p} \pi^{+} \pi^{+}  \pi^{-} [ \Lambda \rightarrow p \pi^{-}] $ decays. Through a combination of integrated pruning and LCAG inference, the HGNN can isolate the beauty-hadron tracks in green and correctly reconstruct the hierarchy of both decay chains.

\subsection{PV association}
\label{sec:pvassociation}
A key additional capability of the HGNN is PV association, which is indicated in the collision event in Figure~\ref{fig:event}, by the blue connections between tracks and PVs. This is most valuable for displaced tracks, such as those arising from beauty-hadron decays and other long-lived particles. Therefore, we quantified the PV association for all tracks in the event, tracks from the decays of beauty hadrons and the PV-association of whole beauty hadrons. As discussed in Section~\ref{sec:architecture}, a given track is associated with a PV by selecting the PV-track edge for that track with the highest last-layer probability score, $\hat{y}^{e_{\rm pv-tr}}_{8}$. Meanwhile, when evaluating the PV association of the HGNN for beauty hadrons we use the ground truth to identify the beauty-hadron decay tracks and associate the beauty hadron with the PV with which the majority of its tracks are associated.

\begin{table}[!h]
\centering
\begin{tabular}{ccc|ccc}
\hline
 \multicolumn{3}{c|}{Method} &  \multicolumn{3}{c}{PV Association (\%)} \\
 \hline
method & task & edge type & track &beauty track & beauty hadron \\
\hline
min IP & - & - & 95.56 ± 0.03 & 88.82 ± 0.21 & 96.14 ± 0.17 \\
MLP & $\mathcal{L}^{\mathrm PV}_{\mathrm BCE}$ & - & 96.37 ± 0.03 & 90.27 ± 0.20 & 96.35 ± 0.17 \\
HGNN & $\mathcal{L}^{\mathrm PV}_{\mathrm BCE}$ & pv-tr & 97.83 ± 0.02 & 96.19 ± 0.14 & 97.60 ± 0.15 \\
HGNN & $\mathcal{L}^{\mathrm PV}_{\mathrm BCE}$ & pv-tr, tr-tr & 99.83 ± 0.01 & 99.74 ± 0.04 & 99.85 ± 0.04 \\
HGNN & $\mathcal{L}^{\mathrm PV}_{\mathrm BCE}, \mathcal{L}^{\mathrm prune}_{\mathrm BCE}, \mathcal{L}^{\mathrm LCA}_{\mathrm CE}$ & pv-tr, tr-tr & 99.88 ± 0.01 & 99.78 ± 0.04 & 99.85 ± 0.04 \\
\end{tabular}
\caption{Comparison of the average PV association (\%) in inclusive beauty-hadron events for various HGNN models with the minimum impact parameter method.}
\label{table:PVassoc}
\end{table}

Table~\ref{table:PVassoc} reports the average PV association accuracy per event for the conventional ``min IP” method, which assigns each track or beauty hadron to the PV with the smallest impact parameter, alongside an MLP baseline and several HGNN variants, distinguished by their training objectives and edge definitions. The MLP, which serves as a non-GNN baseline, is further described in ~\ref{sec:appendixC}. To compute the average PV association accuracy for track / beauty hadron with PV associations, we first calculate, for each event, the PV association accuracy as the ratio of correct associations (numerator) to the total number of associations (denominator). The reported value corresponds to the mean of these per-event accuracies across all events, and the quoted uncertainties represent the standard error of the mean. 

The results show that the HGNN models consistently outperform the conventional min IP approach and MLP baseline across all categories. Incorporating both PV-track and track-track edges yields a substantial improvement, increasing the PV association accuracy for beauty-hadron tracks to around 99.7\%. This demonstrates that relational information between tracks provides a powerful discriminant for identifying the correct PV, effectively resolving ambiguities that the impact-parameter method cannot distinguish. Furthermore, the inclusion of auxiliary tasks, such as the LCA and pruning objectives, yields a small but consistent additional gain, indicating that joint optimization across related tasks improves the network’s ability to model the full event topology.

\begin{figure}[!h]

\begin{center}
\includegraphics[width=8.3cm]{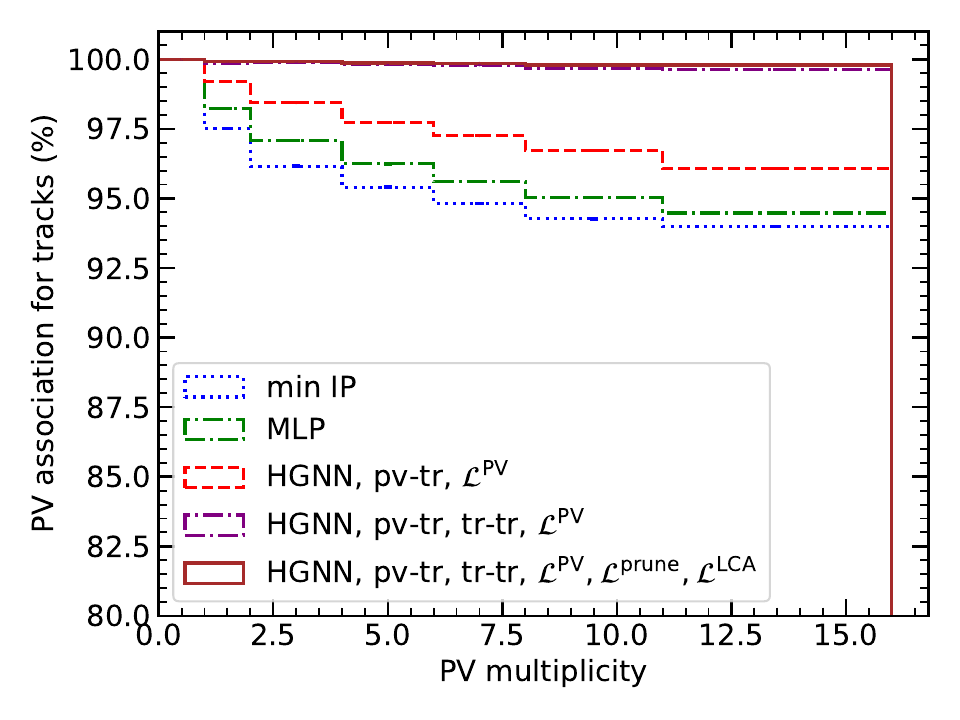}\includegraphics[width=8.3cm]{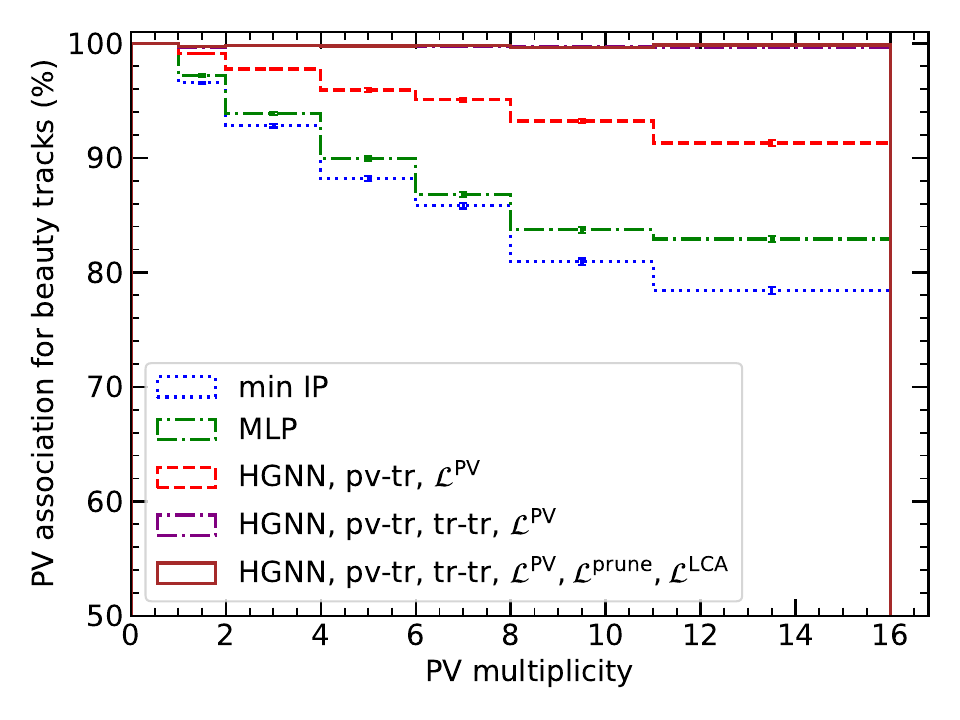}
\caption{Dependence of the average PV-track association (\%) on PV multiplicity for tracks (left) and beauty tracks (right) using various methods for PV association.}
\label{fig:PVassoc}
\end{center}
\end{figure}

\begin{figure}[!h]
\begin{center}
\includegraphics[width=8.3cm]{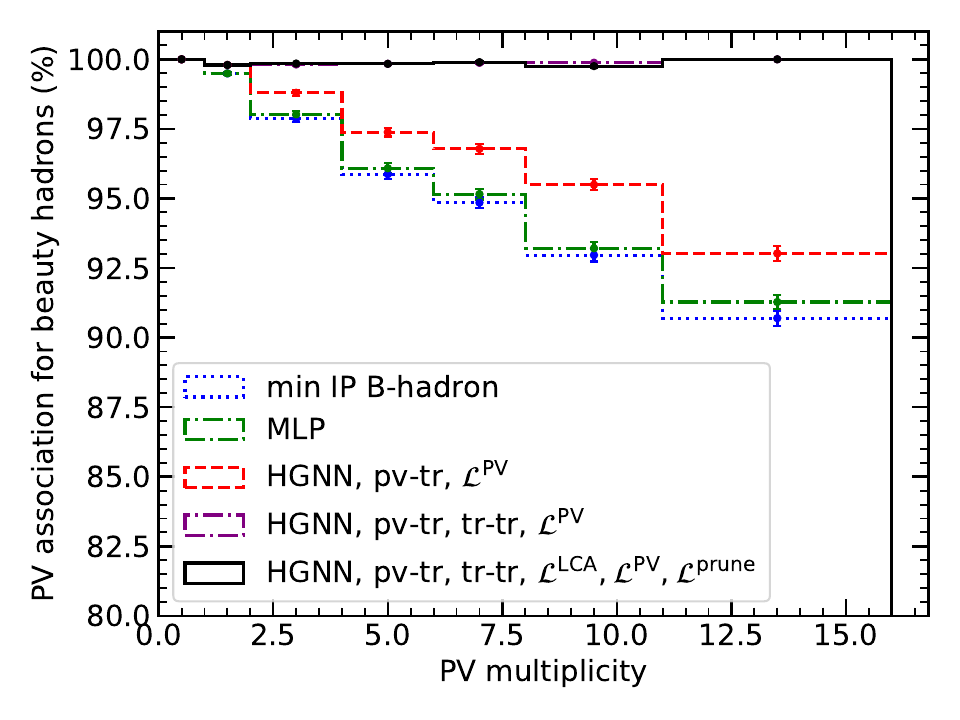}
\caption{Dependence of the average PV association (\%) on PV multiplicity for beauty hadrons using various methods for PV association.}
\label{fig:hadPVassoc}
\end{center}
\end{figure}

Figure~\ref{fig:PVassoc} and Figure~\ref{fig:hadPVassoc} illustrate how the PV association performance of different methods varies with the number of reconstructed PVs. Figure~\ref{fig:PVassoc} shows the dependence of the average PV-track association accuracy for all tracks (left) and beauty-hadron decay tracks (right), while Figure~\ref{fig:hadPVassoc} presents the corresponding results for beauty hadrons. As expected, the association accuracy decreases with increasing PV multiplicity, reflecting the growing ambiguity in assigning tracks or hadrons to the correct PV in more complex events. The conventional min IP method followed by the MLP baseline exhibit the strongest degradation, particularly for beauty-hadron decay tracks, due to their displaced nature arising from the relatively long beauty-hadron lifetime. In contrast, all HGNN variants maintain significantly higher accuracy across the full range of PV multiplicities, demonstrating improved robustness in high-occupancy environments. Among the HGNN configurations, the models incorporating both track-track and PV-track edges show the best overall performance, with the multi-task variant providing a modest additional gain.

%% file: discussion.tex
The ablation studies in Section~\ref{sec:ablation} demonstrate that the inclusion of additional learning objectives noticeably increases the accuracy of the underlying LCAG reconstruction task. This is especially the case for the pruning task, as a high discrimination between the signal and background nodes is essential for correct LCAG reconstruction. Having both track edge and node pruning targets at each layer provides important information, helping models learn richer representations that better discriminate between background and signal nodes. Although smaller in impact, the PV association task also improves the LCAG accuracy, because the tracks of a given beauty hadron are associated with the same PV.

The GNN and HGNN architectures with integrated pruning demonstrate substantially improved reconstruction performance over DFEI for inclusive beauty-hadron decays. This improvement was driven by the effective background isolation achieved via tight pruning selections and improved LCAG classification. Although the initial training of the HGNN (H1 in Table~\ref{table:exclusive}) showed a tendency to prune signal particles in rare exclusive hadronic decays ($\sim$$10^{-5}$ frequency or below~\cite{ParticleDataGroup:2024cfk}), a potential issue for trigger applications, this was effectively addressed. By incorporating a subset of these exclusive decays into the HGNN training process (H2 in Table~\ref{table:exclusive}), perfect reconstruction rates exceeding 90\% were achieved for these challenging modes, with improvements generalising to similar decays. This highlights the importance of tailored training data to optimise performance on rare topologies. The resulting overall gains in performance would have a major impact on both data acquisition and offline analysis, enabling the highly efficient retention of diverse beauty-hadron decays with minimal background.

Another key outcome for practical applications is the acceleration of the inference time provided by integrated pruning. This enables faster CPU/GPU processing for high track multiplicities with minimal performance loss, which is essential for the model's scalability, as the particle collision multiplicity increases under higher luminosity conditions. The proposed weighted message passing scheme, which uses edge and node probability scores, $\hat{y}^{e/v}$, as weights during message passing, allows greater performance invariance when pruning. This can be explained by the learned representations being less dependent on background nodes and edges given their small weighting in message passing.
 
 In addition to providing substantial efficiency gains over DFEI, the HGNN architecture offers the important advantage of enabling precise PV association. The architectural advances proposed in this study have the potential to significantly enhance PV association performance at LHCb. This improvement would, in turn, lead to greater sensitivity across the LHCb physics program by reducing the background contributions and enhancing the resolution of key observables. A more accurate PV association would have a particularly strong impact on the precision of measurements involving decays with missing particles, such as neutrinos, by enabling better determination of the B-hadron flight direction.

%% file: future_work.tex
Building on the successful application of HGNNs to beauty hadron reconstruction, several avenues for future development emerge. A major current limitation is the lack of neutral particles, which is potentially challenging to address, primarily because of their lower reconstruction efficiency and resolution compared to charged particles. Within an HGNN, neutral particles can be represented using their own unique representations and learning tasks, which may help overcome the difficulties associated with their inclusion. For instance, in multi-task learning, a curriculum learning approach could be employed in which the model first focuses on easier tasks, such as learning the LCAG for charged particles and later shifts its attention to tasks involving neutral particles. Beyond incorporating neutrals, several other tasks can be performed to aid reconstruction, such as the reconstruction of secondary decay vertices for track-track edges, flavour tagging of reconstructed beauty hadron decays and particle identification of particle nodes.

Several avenues remain for optimisation. First, we did not conduct a comprehensive joint tuning of the architectural and training hyperparameters. Second, the edge and node pruning thresholds at each HGNN layer were fixed rather than individually optimised to balance the reconstruction accuracy against the inference latency. Third, integrating pruning into the training procedure via a ``pruning-aware” strategy could yield additional performance gains. Finally, complementary techniques, such as model quantisation and weight sparsification, could be applied alongside graph pruning to further accelerate inference.

In this study we used a custom simulation environment from DFEI~\cite{GarciaPardinas:2023pmx}, which mimics the Run 3 environment of LHCb. While the simulation considers several reconstruction effects, which are accounted for through experimentally motivated smearing, its limitations (such as the lack of falsely reconstructed tracks and focus on the vertex detector region) necessitate independent confirmation of the promising performance results using official LHCb simulation and eventually data. Furthermore, additional studies are required to investigate the HGNN reconstruction performance for the high-luminosity conditions of LHCb's Upgrade II. Although higher particle and PV multiplicities pose significant challenges, timing information can be incorporated into the heterogeneous graph representation to retain performance.

%% file: conclusion.tex
Reconstructing beauty hadron decays amid the high-luminosity conditions of current and future LHC runs presents formidable challenges, including higher particle multiplicities and the prevalence of overlapping primary vertices (PVs), further compounded by strict latency and storage requirements. To address these issues, we developed a novel HGNN architecture that jointly prunes background particles, reconstructs beauty-hadron decays, and associates tracks with their correct PV. By integrating pruning at every layer and optimising with a multi-objective loss, our HGNN achieves substantially higher beauty-hadron reconstruction efficiency and background rejection than the earlier DFEI multi-stage GNN framework~\cite{GarciaPardinas:2023pmx}, while limiting the CPU/GPU inference time as the particle multiplicity increases. 

Central to our design are the unique representations for particle tracks and candidate vertices, enabling the HGNN to learn their mutual relations and assign each track to its true origin vertex with high accuracy. Early-layer pruning coupled with weighted message passing reduces the graph size with negligible performance loss, bounding inference latency even as particle multiplicity grows. These architectural innovations deliver substantial gains in PV association and beauty reconstruction performance, while moving closer to the stringent latency and storage requirements for data acquisition.

The potential improvement in PV association and beauty hadron reconstruction would translate into improved sensitivities of precision measurements across the LHCb physics program, reducing sources of background and improving resolutions of derived quantities. More broadly, the architectural novelties are applicable in several particle physics experiments, where heterogeneous datasets, multi-task reconstruction and the scalability of GNN inference time are common.

%% file: acknowledgements.tex
\section{Acknowledgements}

W. S. and N. S. received support from the Swiss National Science
Foundation (SNF) under TMAG-2\_209263. J.G.P. was supported by the U.S. National Science Foundation under Grant No. 2411204. Furthermore, we acknowledge support from the Italian national
funding agency INFN and CERN.

%% file: declaration.tex
\section{Conflicts of Interest}

On behalf of all authors, the corresponding author
states that there is no conflict of interest.

%% file: data_availability.tex
\section{Data availability statement}

The data and code supporting the findings of this study are openly available at the following URLs: \url{https://zenodo.org/records/15584745} and \url{https://github.com/willsutcliffe/scalable_mtl_hgnn}

%% file: appendix.tex
\section{Training curves}
\label{sec:appendixA}
\begin{figure}[!h]
\begin{center}
\includegraphics[width=8.1cm]{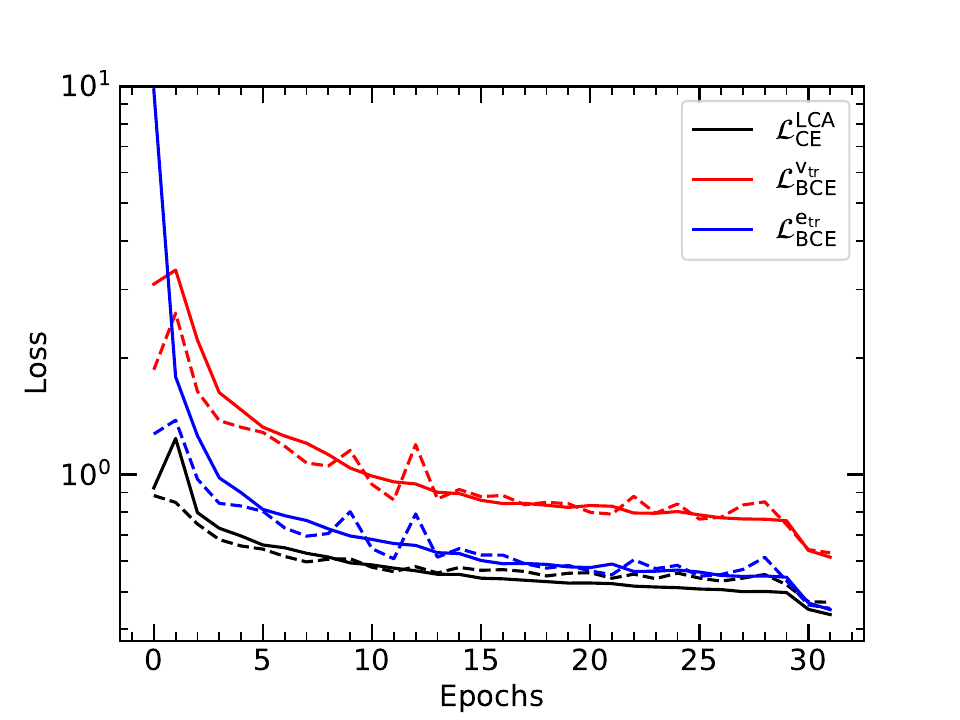}\includegraphics[width=8.1cm]{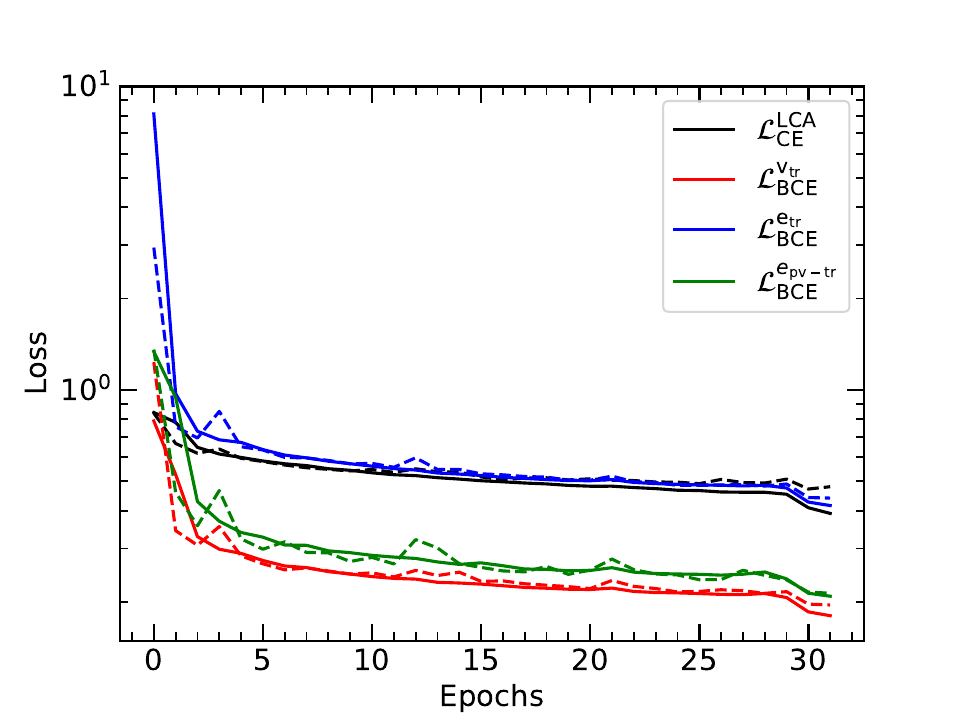}
\caption{Training and validation loss curves with training epochs. The curves are distinguished with solid lines for training and dashed lines for validation.}
\label{loss}
\end{center}
\end{figure}

The multi-task training simultaneously optimises several objectives losses. Figure~\ref{loss} shows the contributions to the overall training and validation loss in Equation~\ref{eq:loss} for the GNN and HGNN trainings as a function of training epochs. 

\section{Multi-task hyperparameter optimisation}

\begin{figure}[!h]
\begin{center}
\includegraphics[width=8.5cm]{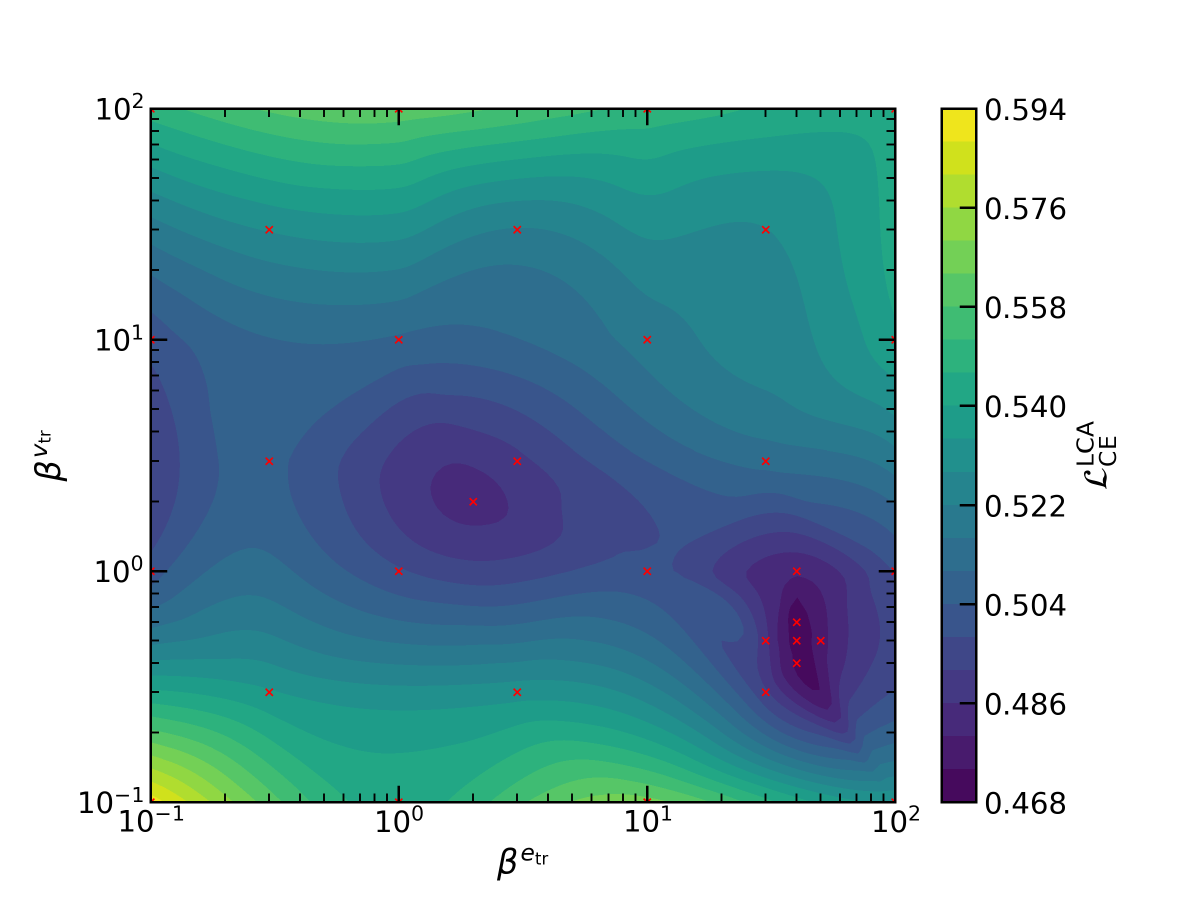}
\end{center}
\caption{Grid scan in $\beta^{e_{\rm tr}}$ and $\beta^{v_{\rm tr}}$ showing the final validation LCAG loss value, $\mathcal{L}^{\rm LCA}_{\rm CE}$.}
\label{fig:gridscan}

\end{figure}

The improvement in performance arising from multi-task training is dependent on the $\beta$ parameters, which govern the significance of the various tasks. To demonstrate this, we performed a two-dimensional grid scan in $\beta^{e_{\rm tr}}$ and $\beta^{v_{\rm tr}}$ for the GNN architecture as shown in Figure~\ref{fig:gridscan}, which indicates that the relative scaling of the edge and node pruning tasks influences the final $\mathcal{L}^{\rm LCA}_{\rm CE}$ value. 

\label{sec:hyper}

Each training was performed in a manner similar to the procedure described in Section~\ref{sec:ablation}. A global minimum is found approximately in the region $(\beta^{e_{\rm tr}}, \beta^{v_{\rm tr}}) = (40, 0.6)$, which indicates that a relatively high weighting of the edge pruning task benefits the LCAG reconstruction task.

\section{MLP for PV association}
\label{sec:appendixC}

The MLP baseline is a feed-forward neural network with five fully connected layers of 128 hidden units and ReLU activations, trained with a binary cross-entropy (BCE) loss to predict whether a given PV-track or PV-beauty-hadron pair corresponds to a correct association.

For PV-track pairs, the input features are equivalent to the node and edge features used by the HGNN. Specifically, for a given PV-track edge, the inputs include the track and PV node features described in Section~\ref{sec:training}, together with the edge feature, namely the impact parameter.

For PV-beauty-hadron pairs, additional feature engineering is applied. The inputs comprise the summed momentum of the tracks originating from the beauty hadron, the decay origin of the beauty hadron, the PV position coordinates, the impact parameter (IP) between the summed momentum of the tracks and the PV, and two boolean indicators specifying whether the PV corresponds to the minimum-IP or second from minumum-IP PV on an event level.

When applied to an event, the network outputs a probability score for the association of each track or beauty hadron to each PV candidate. The associated PV is taken as the one with the highest probability. Training uses the same dataset splits as for the HGNNs.